\documentclass[aps,prd,twocolumn,preprintnumbers,superscriptaddress,nofootinbib,floatfix]{revtex4-1}
\usepackage[dvipsnames]{xcolor}
\usepackage{amssymb,amsmath,amsfonts,bm,slashed,soul,ragged2e,graphicx,epstopdf,hyperref,array}
\usepackage[utf8]{inputenc}
\usepackage{colortbl,color,caption,subcaption}
\enlargethispage{\baselineskip}
\definecolor{steelblue}{RGB}{25,25,112}
\definecolor{dullblue}{rgb}{0,0.298,0.49}
\definecolor{darkred}{rgb}{0.545,0,0}
\definecolor{blue2}{cmyk}{1, 0.1, 0.1, 0}
\hypersetup{colorlinks,linkcolor={darkred},citecolor={blue},urlcolor={dullblue}}
\DeclareCaptionJustification{justified}{\justifying}
\everymath{\displaystyle} 

\begin{document}
\title{Superradiance evolution of black hole shadows revisited}

\author{Rittick Roy}
\email{rittickrr@gmail.com}
\affiliation{Center for Field Theory and Particle Physics and Department of Physics, Fudan University, 200438 Shanghai, P.\ R.\ China}
\author{Sunny Vagnozzi}
\email{sunny.vagnozzi@ast.cam.ac.uk}
\affiliation{Kavli Institute for Cosmology (KICC) and Institute of Astronomy,\\University of Cambridge, Madingley Road, Cambridge CB3 0HA, United Kingdom}
\author{Luca Visinelli}
\email{luca.visinelli@sjtu.edu.cn}
\affiliation{Tsung-Dao Lee Institute (TDLI) and School of Physics and Astronomy, Shanghai Jiao Tong University, 200240 Shanghai, P.\ R.\ China}
\affiliation{INFN, Laboratori Nazionali di Frascati, C.P.\ 13, 100044 Frascati, Italy}

\date{\today}

\begin{abstract}
\noindent Black hole (BH) shadows can be used to probe new physics in the form of ultra-light particles via the phenomenon of superradiant instability. By directly affecting the BH mass and spin, superradiance can lead to a time evolution of the BH shadow, which nonetheless has been argued to be unobservable through Very Long Baseline Interferometry (VLBI) over realistic observation timescales. We revisit the superradiance-induced BH shadow evolution including the competing effects of gas accretion and gravitational wave (GW) emission and, as a first step towards modelling realistic new physics scenarios which predict the existence of multiple ultra-light species, we study the system in the presence of two ultra-light bosons, whose combined effect could help reducing the shadow evolution timescale. We find that accretion and GW emission play a negligible role in our results (justifying previous simplified analyses), and that contrary to our intuition the inclusion of an additional ultra-light boson does not shorten the BH shadow evolution timescale and hence improve detection prospects. However, we point out an important subtlety concerning the observationally meaningful definition of the superradiance-induced BH shadow evolution timescale, which reduces the latter by about an order of magnitude, opening up the possibility of observing the superradiance-induced BH shadow evolution with upcoming VLBI arrays, provided angular resolutions just below the $\mu{\rm as}$ level can be reached. As a concrete example, we show that the angular size of the shadow of SgrA$^*$ can change by up to $0.6\,\mu{\rm as}$ over a period as short as $16$ years, which further strengthens the scientific case for targeting the shadow of SgrA$^*$ with next-generation VLBI arrays.
\end{abstract}

\maketitle

\section{Introduction}
\label{sec:intro}

Black holes (BHs), classical vacuum solutions of virtually any relativistic metric theory of gravity which are believed to constitute the final evolutionary state of sufficiently massive stars, are among the most extreme and peculiar regions of space-time~\cite{Einstein:1916vd,Schwarzschild:1916uq}. We have now entered an age in which BHs and/or observational effects associated to BHs are routinely observed: these effects range from the acceleration of extragalactic cosmic rays~\cite{Anchordoqui:2018qom}, the orbital dynamics of the S2 star around the supermassive BH (SMBH) Sagittarius A$^*$ (SgrA$^*$) at the center of the Milky Way~\cite{GRAVITY:2020gka}, the emission of highly relativistic astrophysical jets by SMBHs at the center of blazars~\cite{Halzen:2002pg}, the signatures of gas accretion by SMBHs in the early Universe or at the center of active galactic nuclei (AGN) and quasars~\cite{Lyke:2020tag}, the detection of gravitational waves (GWs) from the coalescence of BH-BH or BH-neutron star binaries~\cite{LIGOScientific:2016aoc,LIGOScientific:2017vwq}, finally culminating with the first images of the shadow~\cite{EventHorizonTelescope:2019dse,EventHorizonTelescope:2019uob,EventHorizonTelescope:2019jan,EventHorizonTelescope:2019ths,EventHorizonTelescope:2019pgp,EventHorizonTelescope:2019ggy} and magnetic field structure~\cite{EventHorizonTelescope:2021bee, EventHorizonTelescope:2021srq} of the SMBH at the center of the nearby giant elliptical galaxy Messier 87$^*$ (M87$^*$), as captured by the Event Horizon Telescope (EHT) collaboration. This wealth of observations opens up the fascinating prospect of exploiting BHs and their extreme environments as laboratories to test fundamental physics at equally extreme scales, which cannot be reached by terrestrial experiments. One exciting possibility in this sense makes use of the so-called \textit{superradiance} mechanism~\cite{Dicke:1954zz,Penrose:1971uk,1971JETPL..14..180Z,Misner:1972kx,Starobinsky:1973aij}.

Superradiance mechanisms refer to a class of (stimulated or spontaneous) radiation-enhancement phenomena, intimately tied to dissipative systems, and appearing in disparate areas of physics beyond BHs, including optics, quantum mechanics, and condensed matter physics. In the case of BHs, it is the existence of an event horizon (EH) which effectively provides the otherwise dissipationless vacuum with an intrinsic dissipative mechanism. Here, we shall specialize to the case of (rotational) BH superradiance (BHSR), a phenomenon allowing for massive bosonic fields incident upon the EH to be superradiantly amplified and emerge with more energy, at the expense of the BH mass and angular momentum, provided the EH angular velocity is sufficiently high: see Ref.~\cite{Brito:2015oca} for a recent comprehensive review on BHSR. Following the initial seminal works of Refs.~\cite{Dicke:1954zz,Penrose:1971uk,1971JETPL..14..180Z,Misner:1972kx,Starobinsky:1973aij}, the phenomenon of BHSR has been studied in great detail in the literature (see e.g.\ Refs.~\cite{Cardoso:2004nk,Bredberg:2009pv,Cardoso:2012zn,Herdeiro:2013pia,East:2013mfa,Degollado:2013bha,Brito:2014nja,Arvanitaki:2014wva,Rosa:2015hoa,Cardoso:2015zqa,Wang:2015fgp,Rosa:2016bli,East:2017ovw,Cardoso:2017kgn,Rosa:2017ury,Barack:2018yly,Degollado:2018ypf,Ikeda:2018nhb,Ficarra:2018rfu,Baumann:2019eav,Cardoso:2020hca,Dima:2020rzg,Brito:2020lup,Stott:2020gjj,Blas:2020kaa,Mehta:2020kwu,Baryakhtar:2020gao,Unal:2020jiy,Franzin:2021kvj,Caputo:2021efm,Mehta:2021pwf,Cannizzaro:2021zbp,Jiang:2021whw,Karmakar:2021wbs,Khodadi:2021mct} for an inevitably incomplete selection of more recent important works on BHSR). Note that while here we shall focus on superradiance from spin-zero bosons, superradiance from higher spin bosons can and has been considered as well~\cite{Endlich:2016jgc, Baryakhtar:2017ngi, Frolov:2018ezx, Cardoso:2018tly}, see Ref.~\cite{Brito:2015oca} for more details.

If the boson in question is sufficiently heavy, its mass provides a natural potential barrier which confines the field around the BH, allowing it to grow exponentially and rendering the system unstable. This allows for the growth of a boson cloud containing exponentially large population numbers, and provides a realization of Press and Teukolsky's ``BH bomb'' gedanken experiment~\cite{Press:1972zz}. BHSR then effectively results in the formation of a \textit{gravitational atom} featuring hydrogen-like states between the BH and the boson cloud. BHSR can be shown to be most effective when the boson's Compton wavelength is comparable to the BH gravitational radius: in other words, when the product of the boson and BH masses is of order unity in Planck units. From this it follows that BHs with masses in the range ${\cal O}(1) \lesssim M/M_{\odot} \lesssim {\cal O}(10^{10})$ may be used to probe (ultra)light bosons with masses in the range $10^{-22} \lesssim \mu/{\rm eV} \lesssim 10^{-12}$.

Light bosonic fields are ubiquitous in well-motivated extensions of the standard model of particle physics (SM). From the theoretical perspective, scenarios featuring a copious amount of light bosonic fields occur naturally in string theory compactifications (in connection to the ``string axiverse''), wherein the bosons themselves appear as Kaluza-Klein zero-modes of antisymmetric tensor fields~\cite{Witten:1984dg,Svrcek:2006yi,Arvanitaki:2009fg, Cicoli:2012sz,Visinelli:2018utg}. From the phenomenological perspective, light bosonic fields appear frequently in models for the dark matter (DM) and dark energy (DE) which permeate the Universe (see e.g.\ Refs.~\cite{Hu:2000ke,Riotto:2000kh,Schive:2014dra,Hlozek:2014lca,Foot:2014uba,Hui:2016ltb,Irsic:2017yje,Visinelli:2017imh,Hlozek:2017zzf,Rogers:2020ltq} for DM and Refs.~\cite{Wetterich:1987fm,Ratra:1987rm,Caldwell:1997ii,Zlatev:1998tr,Nomura:2000yk,Khoury:2003rn,Cai:2009zp,Burrage:2016bwy,DiValentino:2019exe,Choi:2021aze} for DE). Ultralight bosonic fields also underlie some of the simplest proposed models for early DE, a hypothetical DE-like component invoked to address the Hubble tension~\cite{Poulin:2018cxd,Niedermann:2019olb,Sakstein:2019fmf,Freese:2021rjq,Vagnozzi:2021gjh}. The archetype of several of these models is the axion, a nearly massless pseudo-scalar Goldstone boson associated to the spontaneous breaking of a hypothetical Abelian symmetry~\cite{Peccei:1977hh,Weinberg:1977ma,Wilczek:1977pj,Marsh:2015xka,DiLuzio:2020wdo}.

While BHs themselves cannot strictly speaking be viewed, it is possible to observe their imprint on the surrounding electromagnetic distribution. In particular, a number of seminal studies have shown that a BH surrounded by an optically thin, geometrically thick emission region will appear to a distant observer as a central dark silhouette against the backdrop of a bright ring of finite size: the dark silhouette is typically referred to as the \textit{black hole shadow}~\cite{Luminet:1979nyg,Falcke:1999pj}. In April 2019 the Event Horizon Telescope (EHT), a Very Long Baseline Interferometry (VLBI) radio telescope array with Earth-wide baseline and operating at ${\cal O}({\rm mm})$ wavelengths~\cite{Fish:2016jil}, delivered the first image of the shadow of M87$^*$, the SMBH lurking at the center of the nearby galaxy Messier 87 (M87*)~\cite{EventHorizonTelescope:2019dse,EventHorizonTelescope:2019uob,EventHorizonTelescope:2019jan,EventHorizonTelescope:2019ths,EventHorizonTelescope:2019pgp,EventHorizonTelescope:2019ggy}, with what is easily one of the most iconic images of BH physics.

The vast majority of BHs in the Universe (including M87$^*$ and SgrA$^*$) are low-luminosity AGNs operating at sub-Eddington accretion rates, powered by radiatively inefficient advection-dominated accretion flows (ADAFs) which are, at least at certain observation frequencies, optically thin and geometrically thick~\cite{1998tbha.conf..148N, Yuan:2014gma}. Under these circumstances, the size of the BH shadow reflects the apparent (gravitationally lensed) size of the photon region (the boundary of the region of space-time which allows for the existence of closed spherical photon orbits), and therefore probes the geometry of space-time in the proximity of the EH.~\footnote{The spherical symmetry of the Schwarzshild metric leads to the existence of a photon sphere. Here, we use the terms ``photon region'' and ``spherical photon orbits'' in place of ``photon sphere'' and ``circular photon orbits'' to reflect the fact that the Kerr solution is not spherically symmetric, but only axially symmetric.} For the Kerr-Newman family of BHs, the shadow radius is proportional to the BH mass $M$, and decreases for increasing spin (with the shadow itself becoming asymmetric along the spin axis) and/or electric charge, while approaching the value of $3\sqrt{3}M$~\footnote{More precisely, the shadow radius is $3\sqrt{3}GM$ in the units we have adopted ($3\sqrt{3}GM/c^2$ in SI units). However, as common practice in the literature, throughout the paper when discussing the sizes of BH shadows we shall implicitly set $G=1$.} in the non-rotating, uncharged Schwarzschild limit (see e.g.\ Refs.~\cite{Cunha:2018acu,Dokuchaev:2019jqq,Perlick:2021aok} for recent reviews on theoretical and observational aspects of BH shadows). It was recently demonstrated in Refs.~\cite{Narayan:2019imo,Volkel:2020xlc,Bronzwaer:2021lzo,Ozel:2021ayr,Younsi:2021dxe} that, under the ADAF assumption, the resulting size and morphology of the BH shadow is to a large extent insensitive to the details of the surrounding accretion flow, addressing concerns raised earlier in Refs.~\cite{Gralla:2019xty,Gralla:2019drh,Gralla:2020yvo,Gralla:2020srx,Gralla:2020pra}.\footnote{An interesting analogy in this sense, first pointed out in Ref.~\cite{Bronzwaer:2021lzo}, views this as being similar to the resulting impression of a coin placed on a white paper which is then rubbed with a crayon, with the resulting shape of the coin impression being independent of crayon-dependent factors.} This simple but crucially important aspect implies that, once the BH's mass-to-distance ratio is known, BH shadows may be used to test fundamental physics, as has been done in several recent works focusing on the shadow of M87$^*$ imaged by the EHT (see e.g. Refs.~\cite{Moffat:2019uxp,Giddings:2019jwy,Wei:2019pjf,Shaikh:2019fpu,Tamburini:2019vrf,Konoplya:2019nzp,Contreras:2019nih,Bar:2019pnz,Jusufi:2019nrn,Vagnozzi:2019apd,Banerjee:2019cjk,Long:2019nox,Zhu:2019ura,Contreras:2019cmf,Qi:2019zdk,Neves:2019lio,Javed:2019rrg,Tian:2019yhn,Banerjee:2019nnj,Shaikh:2019hbm,Kumar:2019pjp,Allahyari:2019jqz,Li:2019lsm,Jusufi:2019ltj,Rummel:2019ads,Kumar:2020hgm,Li:2020drn,Narang:2020bgo,Liu:2020ola,Konoplya:2020bxa,Guo:2020zmf,Pantig:2020uhp,Wei:2020ght,Kumar:2020owy,Islam:2020xmy,Chen:2020aix,Jin:2020emq,Sau:2020xau,Jusufi:2020dhz,Kumar:2020oqp,Chen:2020qyp,Zeng:2020dco,Neves:2020doc,Ovgun:2020gjz,Badia:2020pnh,Jusufi:2020cpn,Khodadi:2020jij,Belhaj:2020nqy,Jusufi:2020agr,Kumar:2020yem,Jusufi:2020odz,Belhaj:2020mlv,Kruglov:2020tes,Saurabh:2020zqg,EventHorizonTelescope:2020qrl,Contreras:2020kgy,Ghasemi-Nodehi:2020oiz,Ghosh:2020spb,Khodadi:2020gns,Lee:2021sws,Contreras:2021yxe,Shaikh:2021yux,Afrin:2021imp,Addazi:2021pty,EventHorizonTelescope:2021dqv,Shaikh:2021cvl,Badia:2021kpk,Wang:2021irh,Khodadi:2021gbc,Cai:2021uov,Wei:2021lku,Rahaman:2021web,Walia:2021emv,Ghasemi-Nodehi:2021ipd,Afrin:2021wlj,Jusufi:2021fek,Rahaman:2021kge,Cimdiker:2021cpz,Pal:2021nul,Jha:2021bue}).

The dependence of the sizes of BH shadows on the BH mass and spin unlocks the possibility of observing the imprint of BHSR on the shadow, as BHSR ``feeds'' the boson cloud by spinning down BHs and decreasing their masses. This is expected to lead to an \textit{evolution} of BH shadows as the BHSR process goes on. The possibility of observing this shadow evolution and its use to constrain light bosons has been studied earlier, including by one of us, in Refs.~\cite{Roy:2019esk,Creci:2020mfg} (see also Refs.~\cite{Davoudiasl:2019nlo,Cunha:2019ikd} for related works), although these works pointed out that appreciable changes due to this evolution occur on timescales which are at best of ${\cal O}(10^2){\rm\, yrs}$ (but in most cases longer), way longer than realistic observation times.

In this work, we are nevertheless prompted to revisit the question of whether the BHSR-induced shadow evolution is (un)observable. In particular, we go beyond the earlier Refs.~\cite{Roy:2019esk,Creci:2020mfg} in various respects. First of all, several BHSR studies are based on a linearized analysis which neglects competing effects whose impact on the development on the process can be non-negligible. Two key effects in this respect are (gas) accretion and gravitational wave (GW) emission, whose effect we include in our analysis. Another crucial point is that several realistic scenarios that extend the SM predict the existence of more than one (ultra-)light boson: the string axiverse is a well-studied example in this sense. This observation motivates us to revisit the BHSR-induced shadow evolution in the presence of multiple light bosons, as the combined effect thereof might speed up the BHSR process and therefore reduce the BH shadow evolution timescale. While we find that our expectation is actually not met, we also point out a subtlety in the definition of the ``meaningful'' BH shadow evolution timescale: taking this into account reduces the evolution timescale by about an order of magnitude, thereby enabling the possibility of observing the BHSR-induced shadow evolution with upcoming space-based VLBI arrays.

The rest of this paper is then organized as follows. In Sec.~\ref{sec:superradianceshadows}, we review the BH superradiance process (including competing effects such as gas accretion and GW emission) and the computation of BH shadows from a more quantitative standpoint. In Sec.~\ref{sec:methodology} we discuss our methodology, including the numerical system we solve and the observables we consider. Our results are reported in Sec.~\ref{sec:results} and discussed in Sec.~\ref{sec:discussion}. Finally, in Sec.~\ref{sec:conclusions} we provide concluding remarks. Throughout this paper we shall adopt natural units with $\hbar=c=1$.

\section{Black hole superradiance and black hole shadows}
\label{sec:superradianceshadows}

In this Section, we fix the notation used for the Kerr metric in Sec.~\ref{subsec:kerrmetric} and the boson field Lagrangian in Sec.~\ref{subsec:bosonfield}, we review the mechanism of BH superradiance (BHSR) in Sec.~\ref{subsec:bhsuperradiance}, our treatment of gas accretion and GW emission in Sec.~\ref{subsec:accretionemission}, and the computation of BH shadows in Sec.~\ref{subsec:bhshadows}.

\subsection{Kerr metric}
\label{subsec:kerrmetric}

The space-time around a rotating BH of mass $M$ and carrying angular momentum of magnitude $J$ is described by the Kerr metric. In Boyer–Lindquist coordinates, the line element of the Kerr metric is given by:
\begin{eqnarray}
\mathrm{d}s^2 &=& -\mathrm{d}t^2 + \frac{2r_gr}{\Sigma}\left(\mathrm{d}t - a\sin^2\theta\mathrm{d}\varphi\right)^2 + \frac{\Sigma}{\Delta}\mathrm{d}r^2 + \Sigma\mathrm{d}\theta^2 +\nonumber\\
&& + \left(r^2+a^2\right)\sin^2\theta\mathrm{d}\varphi^2\,,
\label{eq:kerrmetric}
\end{eqnarray}
where $t$ is cosmic time as measured far from the BH, $(r,\theta,\varphi)$ are spherical coordinates centered around the BH, $r_g = GM$ is the BH gravitational radius, the spin per unit mass $a \equiv J/M$ has units of length, and where:
\begin{equation}
\Sigma = r^2+a^2\cos^2\theta\,,\quad \Delta= r^2-2r_gr+a^2\,.
\end{equation}
The Kerr metric reduces to the Schwarzschild metric describing a non-rotating BH in the limit $a\to 0$, as the term $\propto \mathrm{d}t\mathrm{d}\varphi$ disappears. The event horizon for a Kerr BH is determined by setting $\Delta = 0$, in other words:
\begin{equation}
\label{eq:rhorizon}
r_\pm = r_g \pm \sqrt{r_g^2-a^2}\,,
\end{equation}
so that the event horizon exists only for $a \leq r_g$, which is known as the Kerr bound. It is conjectured that all rotating BHs follow such a condition, or equivalently $a_{\star} \leq 1$, where $a_{\star}\equiv a/r_g = J/(GM^2)$ is the dimensionless spin, so that the BH singularity is protected by a horizon.~\footnote{See Refs.~\cite{Gimon:2007ur, Bambi:2008jg} for a discussion of ``superspinar'' solutions that evade the Kerr bound.}

We focus on the region between the outer horizon $r_H \equiv r_+$ and the static limit horizon, defined as the surface for which the argument of $\mathrm{d}t$ vanishes,
\begin{equation}
r_t = r_g + \sqrt{r_g^2-a^2\cos^2\theta}\,.
\end{equation}
In the region $r_H \leq r \leq r_t$, a test object cannot remain stationary and would be forced to spin along with the BH. Particles falling into this region right outside of the horizon, called the ergosphere, can gain energy from the spinning BH and escape from it. The existence of the ergosphere allows for the extraction of rotational energy from the spinning BH in the form of energetic particles, through a process known as the Penrose BHSR mechanism~\cite{Penrose:1971uk}. We discuss the energy extraction in Sec.~\ref{subsec:bhsuperradiance}.

\subsection{Massive bosonic fields}
\label{subsec:bosonfield}

Throughout this paper, we shall consider a Kerr-Klein-Gordon system consisting of 2 real (pseudo-)scalar fields $\phi_i$, with action given by:
\begin{equation}
\label{eq:kerrkleingordon}
S \!=\! \int d^4x\sqrt{-g}\, \left ( \frac{R}{16\pi G} -\frac{1}{2}\nabla_{\mu}\phi_i\nabla^{\mu}\phi_i - \frac{1}{2}\mu_i^2\phi_i^2 \right )\,,
\end{equation}
where $g_{\mu\nu}$ is the Kerr metric in Eq.~(\ref{eq:kerrmetric}), $\nabla_{\mu}$ denotes the covariant derivative on the Kerr background, and $\mu_i$ are the masses of the fields $\phi_i$ (with summation over the species index $i=1,2$ implied).

In writing the action in Eq.~(\ref{eq:kerrkleingordon}), we have made a few simplifying assumptions. First of all, we have neglected any effect due to the presence of a cosmological constant, as we are not interested in the cosmological evolution of the system.\footnote{We note that for sufficiently distant systems (e.g.\ SMBHs in the Hubble flow) and for sufficiently long evolution timescales, the cosmological evolution may become relevant~\cite{Perlick:2018iye,Bisnovatyi-Kogan:2018vxl,Tsupko:2019pzg,Tsupko:2019mfo,Vagnozzi:2020quf,Roy:2020dyy,Frion:2021jse}.} Next, we have assumed that the two fields do not mix, i.e.\ that the mass term is diagonal in field space and there is no kinetic mixing. In addition, we have not included self-interaction potentials (e.g.\ a quartic term) for the scalar fields. We have adopted these assumptions in the interest of simplicity, in order to more clearly understand the physical effects driving the superradiant instability and therefore the evolution of the BH shadow, and also given that most of the work on BHSR is framed within the above action. Finally, while for our subsequent computations we work under the assumption of a Kerr background, the action in Eq.~(\ref{eq:kerrkleingordon}) is much more general and our results can be extended rather straightforwardly at least to the Kerr-Newman family of metrics. A more complete exploration of the impact of relaxing these assumptions is deferred to follow-up work.

The Kerr-Klein-Gordon equations describing the full non-linear evolution of the system are
\begin{align}
\left ( \Box -\mu_i^2 \right )\phi_i = 0\,, \label{eq:kleingordon}\\
R^{\mu\nu} \!-\! \frac{g^{\mu\nu}}{2}R = 8\pi G T^{\mu\nu}\,,\label{eq:einstein}\\
T^{\mu\nu} = \nabla^{\mu}\phi_i\nabla^{\nu}\phi_i \!-\! \frac{g^{\mu\nu}}{2}\! \left[ \nabla^{\alpha}\phi_i\nabla_{\alpha}\phi_i \!+\! \mu_i^2\phi^2 \right],\label{eq:tensor}
\end{align}
where $T^{\mu\nu}$ is the stress-energy tensor of the scalar field, $\Box = g^{\mu\nu}\nabla_\mu\nabla_\nu$, $R_{\mu\nu}$ is the Ricci tensor and $R$ is the Ricci scalar. In the following, we will treat the Kerr metric as a background over which the scalar fields propagate, and we assume that backreaction of the fields on the metric is negligible. Later in Sec.~\ref{subsec:backreaction}, we will argue that this is a safe approximation.

\subsection{Black hole superradiance}
\label{subsec:bhsuperradiance}

Incident low-frequency and monochromatic radiation of energy $\omega$ scattering off a spinning BH of mass $M$ and angular momentum $J$ gets superradiantly amplified at the expense of the BH rotational energy, provided the following relation is satisfied:
\begin{equation}
\label{eq:superradiancerelation}
\omega < m\,\omega_+\,,
\end{equation}
where $m$ is the azimuthal quantum number of the unstable mode with respect to the BH rotation axis, and $\omega_+$ is the BH event horizon's angular velocity, which is given in terms of the dimensionless spin $a_{\star}\equiv a/r_g = J/(GM^2)$ by the following expression:
\begin{equation}
\label{eq:omegaplus}
\omega_+ = \frac{a_{\star}}{2r_g \left ( 1+\sqrt{1-a_{\star}^2} \right )}\,.
\end{equation}
Note, as already pointed out earlier, that the Kerr bound implies $0 \leq \vert a_{\star} \vert <1$.

Superradiance results in the BH spinning down and losing mass and angular momentum until Eq.~(\ref{eq:superradiancerelation}) is no longer satisfied. Considering the case where the incident radiation is due to ultralight (pseudo-)scalar particles with mass $\mu$, the Kerr-Klein-Gordon system admits quasi-bound states with complex eigenfrequencies $\omega_{nlm}$, where $n$, $l$, and $m$ are the principal quantum number, angular momentum quantum number, and magnetic/azimuthal quantum number indexing the quantum levels of the corresponding bound states, analogously to the quantum numbers characterizing solutions to the Schr\"{o}dinger equation for the hydrogen atom. Modes which satisfy $\omega_{nlm}<\mu$ are naturally confined by the potential barrier provided by the (pseudo-)scalar field mass, and can grow exponentially over time. The BH will then form large quasi-stationary gravitational bound states (the so-called \textit{gravitational atom} or \textit{gravatom}) with a boson cloud resulting from exponential amplification of a small initial boson population.

It is useful to define the gravitational coupling (which we shall occasionally refer to as mass parameter) $\alpha$ as the ratio between the BH gravitational radius $r_g$ and the light boson's Compton wavelength $\lambda_C$ (equal to $\hbar/(\mu c)$ in SI units). Inserting numbers, the gravitational coupling is approximately given by:
\begin{eqnarray}
\alpha \equiv \frac{r_g}{\lambda_C} \simeq 10^{10} \left ( \frac{M}{M_{\odot}} \right ) \left ( \frac{\mu}{{\rm eV}} \right ) \,.
\label{eq:alpha}
\end{eqnarray}
The superradiant instability is largest when $\alpha \sim 0.42$, and it is generally very efficient for $\alpha \sim {\cal O}(0.1)$~\cite{Dolan:2007mj,Witek:2012tr}. From Eq.~(\ref{eq:alpha}) it therefore follows that the observed BHs with masses in the range ${\cal O}(1) \lesssim M/M_{\odot} \lesssim {\cal O}(10^{10})$ can potentially be used to probe ultralight bosons whose masses span the range $10^{-22} \lesssim \mu/{\rm eV} \lesssim 10^{-12}$.

Let us now discuss in more detail the equations governing the BHSR process. Consider a system of $i$ non-interacting boson species (in our case $i=2$). Quantum levels of the gravatom are indexed by quantum numbers $nlm$ as with the hydrogen atom as discussed above. Here, we shall be concerned with the $n=2,l=1,m=1$ superradiance growth mode, as it is the fastest growing mode~\cite{Arvanitaki:2014wva,Brito:2015oca}, and thus the one leading to the most optimistic results. The evolution of the $i$th boson occupation number for this quantum level, denoted by $N_i$, is governed by the following equation:
\begin{eqnarray}
\label{eq:dotN}
\dot N_i = \Gamma_i\,N_i\,,
\end{eqnarray}
where the growth rate $\Gamma_i$ is given in terms of the horizon radius $r_+$ in Eq.~(\ref{eq:rhorizon}) as~\cite{Detweiler:1980uk,Baryakhtar:2017ngi}:
\begin{equation}
\label{eq:gammai}
\Gamma_i=\frac{\left ( GM\mu_i \right ) ^9}{24GM} \left ( a_{\star}-2\mu_i r_+ \right )\,.
\end{equation}
Superradiance extracts energy and angular momentum from the spinning BH, effectively slowing the BH down, while a boson cloud of mass $M_i = \mu_i\,N_i$ forms. Energy extraction continues as long as $\Gamma_i>0$ or, equivalently, as long as the dimensionless spin is larger than a minimum spin $a^{(i)}_{\min}$, which from Eq.~(\ref{eq:gammai}) is given by:
\begin{eqnarray}
a^{(i)}_{\min} &=& \frac{2k_i}{1+k_i^2}\,, \label{eq:amin}\\
k_i &=& 2\mu_i M \left ( 1-\frac{\mu_i^2M^2}{2} \right ) ^{-1}\,.\label{eq:ki}
\end{eqnarray}
In the presence of multiple light bosons with masses $\mu_i$ (ordered with increasing mass, i.e.\ $\mu_1$ is the mass of the lightest boson, and so on), we can of course define a minimum spin associated to the action of the $i$th boson, $a^{(i)}_{\min}$. During the superradiance phase, and neglecting the competing effects we will discuss later in Sec.~\ref{subsec:accretionemission}, the mass and angular momentum of the BH change over time with rates given by:
\begin{eqnarray}
\dot{M} &=& -\sum_i \mu_i\dot N_i \,,\label{eq:dotMfinal}\\
\dot{J} &=& -\sum_i j\dot N_i\,, \label{eq:dotJfinal}
\end{eqnarray}
where $j=\sqrt{2}$ is the angular momentum per unit boson mass, the sum runs over the $i$ boson species, and a dot denotes a derivative with respect to $t$.

\subsection{Gas accretion and gravitational wave emission}
\label{subsec:accretionemission}

Although we will neglect backreaction of the scalar cloud on the background geometry as argued later in Sec.~\ref{subsec:backreaction}, there are potentially other important competing effects to be taken into account when discussing BHSR, and which are usually neglected in most (linearized) BHSR analyses in the literature: these are gas accretion and GW emission. In this work we shall consistently take accretion and GW emission into account.

Realistic astrophysical BHs are not isolated systems, but live in complex astrophysical environments, surrounded by matter fields which continuously accrete onto the BHs themselves. There are various ways by means of which the interplay between gas accretion and BHSR can have important consequences for the latter. By increasing the BH's mass and angular momentum, accretion na\"{i}vely \textit{competes} with superradiant extraction which instead acts to decrease these quantities. This is true for a BH for which the superradiant instability phase has already begun. Even then, if superradiant extraction is weak due to the BH not being sufficiently massive and hence $\alpha$ being low (at fixed boson mass $\mu$), accretion can increase the BH mass, which in turn strengthens the superradiance process. Finally, in a different regime, accretion can act as a \textit{trigger} for BHSR, by spinning up BHs whose spin is below $a_{\min}$ and would not otherwise have entered the superradiant instability phase.

Following Ref.~\cite{Brito:2014wla}, we adopt an extremely conservative assumption wherein mass accretion at a rate $\dot{M}_{\rm ACC}$ occurs at a fixed fraction $f_{\rm Edd}$ of the Eddington rate $\dot{M}_{\rm Edd}$ (see for instance Ref.~\cite{Barausse:2014tra}):
\begin{eqnarray}
\dot{M}_{\rm ACC} = f_{\rm Edd}\dot{M}_{\rm Edd} \simeq 0.02f_{\rm Edd}\frac{M(t)}{10^6M_{\odot}}M_{\odot}\,{\rm yr}^{-1},
\label{eq:dotmacc}
\end{eqnarray}
where $M(t)$ is the mass of the BH at time $t$. In our modeling, the BH mass changes over time due to \textit{i)} superradiance, \textit{ii)} accretion, and \textit{iii)} GW emission, as clearly shown in Eq.~\eqref{eq:dotM} below. The estimate in Eq.~\eqref{eq:dotmacc} assumes a radiative efficiency $\eta \approx 0.1$, as required by the comparison between the mass function of BHs and the luminosities of AGNs, see e.g.\ Ref~\cite{Brito:2014wla}. The specific value of the Eddington rate for mass accretion depends on the details of the accretion disk surrounding the BH, and can take a wide range of values from ${\cal O}(1)$ for AGNs to ${\cal O}(10^{-9})$ for quiescent galactic nuclei. Here, we fix the value of the Eddington rate for mass accretion to $f_{\rm Edd}=0.2$. This is somewhat of a conservative estimate~\cite{Brito:2014wla}, and one which in any case does not affect our results, since we later find a posteriori that even for values of $f_{\rm Edd}$ of order unity, the superradiance rate always exceeds the accretion rate by several orders of magnitude.

To estimate the angular momentum accretion rate $\dot{J}_{\rm ACC}$, we again follow Ref.~\cite{Brito:2014wla} and make the conservative assumption that the accretion disk lies on the equatorial plane, and extends down to the innermost stable (prograde) circular orbit (ISCO). If radiation effects are ignored, the angular momentum accretion rate can be expressed as~\cite{Bardeen:1970zz}:
\begin{eqnarray}
\dot{J}_{\rm ACC} &=& \frac{L(M,J)}{E(M,J)}\dot M_{\rm ACC}\,,\label{eq:dotjacc}\\
L(M,J) &=& \frac{2M}{3\sqrt{3}}\left(1+2\sqrt{\frac{3 r_{\rm ISCO}(M,J)}{M} - 2}\right)\,, \label{eq:L}\\
E(M,J) &=& \sqrt{1-\frac{2M}{3 r_{\rm ISCO}(M,J)}}\,, \label{eq:E}
\end{eqnarray}
where $L(M,J)$ and $E(M,J)$ are the angular momentum and energy per unit mass at the ISCO radius $r_{\rm ISCO}$ respectively, and the ISCO location in Boyer-Lindquist coordinates is given in closed form as:
\begin{eqnarray}
r_{\rm ISCO} &=& M\left(3-\sqrt{(3-Z_1)(3+Z_1+2Z_2)}+Z_2\right)\,\label{eq:risco},\\
Z_1 &=& 1+\sqrt[3]{1-a_{\star}^2}(\sqrt[3]{1+a_{\star}}+\sqrt[3]{1-a_{\star}})\,\label{eq:Z1}\,,\\
Z_2 &=& \sqrt{3a_{\star}^2+Z_1^2}\,.\label{eq:Z2}
\end{eqnarray}
Obviously, all three quantities $L$, $E$, and $r_{\rm ISCO}$ are time-dependent through their dependence on $M$ and $J$.

Besides gas accretion, another important effect potentially competing with BHSR is GW emission. In fact, a real scalar cloud will eventually be re-absorbed by the BH and dissipated through GW emission~\cite{Arvanitaki:2010sy,Witek:2012tr,Cardoso:2013krh,Okawa:2014nda}. In general, a monochromatic scalar cloud will (incoherently) emit GWs at a frequency $\lambda \sim \pi/\mu$, and at a rate $P_{\rm GW}$ for which an upper limit is given by (see e.g.\ Refs.~\cite{Yoshino:2013ofa,Brito:2014wla}):
\begin{eqnarray}
P_{\rm GW} = \frac{484+9\pi^2}{23040}\,P_{\rm Pl} \left ( \frac{\mu^2 N^2}{M^2} \right ) \left ( GM\mu \right ) ^{14}\,,
\label{eq:dotEGW}
\end{eqnarray}
where $N$ is the relevant occupation number and $P_{\rm Pl}$ is given in terms of the Planck mass $M_{\rm Pl} = G^{-1/2}$ and the Planck time $t_{\rm Pl} = G^{1/2}$ as $P_{\rm Pl} \equiv M_{\rm Pl}/t_{\rm Pl}\approx 3.63\times10^{52}{\rm\, W}$. In this work, we shall make the conservative assumption of modelling the GW emission rate through Eq.~(\ref{eq:dotEGW}), despite the latter being an upper limit to the true rate. We find a posteriori that the impact of GW emission on our results is anyhow negligible, justifying our approximation. Analogously, we will make the assumption that angular momentum is lost to GW emission at a rate $\dot{J}_{\rm GW}$ given by:
\begin{eqnarray}
\dot{J}_{\rm GW} = \frac{P_{\rm GW}}{\mu}\,.
\label{eq:dotjgw}
\end{eqnarray}

Summing all contributions, the final set of coupled differential equations governing the evolution of the BH mass and angular momentum is given by:
\begin{align}
\dot{M} = -\sum_i \left ( \mu_i\dot N_i+ P_{{\rm GW},i} \right ) + \dot M_{\rm ACC} \,,
\label{eq:dotM}\\
\dot{J} = -\sum_i \left( j\dot N_i+\frac{1}{\mu_i}\,P_{{\rm GW},i} \right) + \frac{L(M,J)}{E(M,J)}\dot M_{\rm ACC}\,.
\label{eq:dotJ}
\end{align}
The above system is the one we will solve throughout the work, although for numerical convenience we will rescale the set of equations as discussed in Sec.~\ref{sec:methodology}.

\subsection{Black hole shadows}
\label{subsec:bhshadows}

A BH surrounded by an emission region which is (at least within a certain wavelength range) optically thin and geometrically thick appears to a distant observer as a central dark region surrounded by a bright ring. The central dark region is referred to as the ``BH shadow'', and is the gravitationally lensed image of the photon region, i.e.\ the region around the BH which supports the existence of closed, spherical photon orbits. In other words, the edge of the BH shadow is a closed curve given by the locus of rays that can escape bound orbits and travel to a distant observer. Hence, the BH shadow edge separates capture and scattering orbits. VLBI arrays can be used to image BH shadows, as the EHT collaboration beautifully demonstrated in 2019 by revealing the image of M87$^*$'s shadow~\cite{EventHorizonTelescope:2019dse,EventHorizonTelescope:2019uob,EventHorizonTelescope:2019jan,EventHorizonTelescope:2019ths,EventHorizonTelescope:2019pgp,EventHorizonTelescope:2019ggy,EventHorizonTelescope:2021bee,EventHorizonTelescope:2021srq}.

As discussed earlier, here we shall focus on Kerr BHs, i.e.\ rotating chargeless BHs, although our results can easily be extended to the more generic family of Kerr-Newman BHs. Kerr BH shadows can be computed by considering null geodesics and focusing on those characterized by unstable photon orbits. Here we shall briefly review this standard procedure, and refer the reader to Refs.~\cite{Cunha:2018acu,Dokuchaev:2019jqq,Perlick:2021aok} for recent reviews on BH shadows and various approaches towards calculating them. In Boyer-Lindquist coordinates, radial null geodesics on a Kerr background are described by the following equation:
\begin{eqnarray}
\label{eq:radial}
(r^2+a^2\cos^2\theta)^2 \left ( \frac{\mathrm{d}r}{\mathrm{d}\lambda} \right ) ^2={\cal R}(r)\,,
\end{eqnarray}
where $\theta$ is the polar angle, and the quantity ${\cal R}$ is characterized by three constants of motion: energy $E=-p_{\rm t}$, angular momentum component parallel to the BH spin $L_z=p_\phi$, and the Carter constant ${\cal Q}$. Motion for a test particle around the BH is only possible for ${\cal R}(r) \geq 0$. As photon trajectories are independent of their energies, Eq.~(\ref{eq:radial}) can be rescaled by dividing it through by a factor $E^2$ and re-expressing it in terms of the quantities $\xi=L_z/E$ and $\eta={\cal Q}/E^2$ (which are also constants of motion), with a new affine parameter $\tilde\lambda=E\lambda$.

The apparent image of the BH's photon region, and therefore the edge of the BH shadow, is governed by unstable spherical photon orbits, separating capture and scattering orbits. These are determined by the conditions:
\begin{eqnarray}
{\cal R}(r_{\star})=0\,, \quad \frac{\partial{\cal R}}{\partial r}(r_{\star})=0\,, \quad \frac{\partial^2{\cal R}}{\partial r^2}(r_{\star}) \geq 0\,,
\label{eq:unstablecircularorbit}
\end{eqnarray}
where $r_{\star}$ is the largest among the real roots of ${\cal R}$. The system given in Eq.~(\ref{eq:unstablecircularorbit}) can be solved to obtain a pair of ``critical'' values for the two constants of motion $(\xi_c,\eta_c)$. The BH shadow as seen by a distant observer can then be obtained by projecting $(\xi_c\,,\eta_c)$ to the celestial coordinates $(x\,,y)$ of an observer situated at infinity:
\begin{eqnarray}
x=\frac{\xi_c}{\sin \iota}\,, \quad y = \pm\sqrt{\eta_c+a^2\cos^2\iota-\xi_c^2\cot^2\iota}\,,
\label{eq:xietaxy}
\end{eqnarray}
where $\iota$ is the observation angle, with $\iota=90^{\circ}$ and $\iota=0^{\circ}$ denoting an observer on the equatorial plane (edge-on) or aligned with the BH spin axis (face-on), respectively. Relativistic effects are largest the more $\iota$ approaches $90^{\circ}$, and result in the BH shadow being asymmetric along the spin axis, in particular being flattened on the side associated to photons whose angular momentum is aligned with the BH spin, due to frame dragging effects.

For ease of comparison to observations, it is convenient to compress the information contained within the shape of the BH shadow in a few relevant observables. This compression practice has been adopted in a number of works, including some by the EHT collaboration itself. We define $x_{\max}$ and $x_{\min}$ to be the maximum values of the celestial $x$ coordinates for the shadow edge, and similarly for $y_{\max}$ and $y_{\min}$. Then, the shadow diameter $R$ is naturally defined as:
\begin{eqnarray}
R = x_{\max}-x_{\min} \equiv \Delta x\,,
\label{eq:diameter}
\end{eqnarray}
and we recall that $R = 6\sqrt{3}M$ for a Schwarzschild BH. Additionally, we define the quantity $\chi$ as follows:
\begin{eqnarray}
\chi = \frac{y_{\max}-y_{\min}}{x_{\max}-x_{\min}} \equiv \frac{\Delta y}{\Delta x}\,.
\label{eq:chi}
\end{eqnarray}
For a Schwarzschild ($a_{\star}=0$) BH, $\chi=1$, reflecting the fact that the BH shadow is a perfect circle. Relativistic effects playing an important role with increasing spin $a_{\star}$ and/or observation angle $\iota$, and resulting in the shadow boundary deviating from a perfect circle, lead to $\chi \geq 1$. In the literature, $\chi$ is usually referred to as ``axis ratio'' or ``oblateness''. Here we shall adopt the former label.

Both $R$ and $\chi$ are useful ``summary observables'' in comparing theory to observations of BH shadows. If one can reliably determine the BH mass (for instance through gravitational lensing, stellar/gas dynamics, or reverberation mapping), or more precisely the BH distance-to-mass-ratio, then $R$ can be directly related to the angular size of the dark region appearing in VLBI observations. Note that $R$ carries units of mass (as stated earlier, we will neglect units of $G$ in this context). The morphology of the dark region can also be used to set limits on $\chi$: for instance, the EHT collaboration reports that $\chi \lesssim 4/3$ for M87$^*$~\cite{EventHorizonTelescope:2019dse}. To sum up, we note that $\chi$ is closely related to other measures of the oblateness of the BH shadow, such as the deviation from circularity $\Delta C$, see e.g.\ Eq.~(5) in Ref.~\cite{Bambi:2019tjh}, constrained by the EHT collaboration to be $\Delta C \lesssim 10\%$ for the shadow of M87$^*$~\cite{EventHorizonTelescope:2019dse}.

\subsection{Backreaction}
\label{subsec:backreaction}

Throughout this work, we will treat the Kerr metric as a background over which the scalar fields propagate, assuming that backreaction of the fields on the metric is negligible. To put it differently, we are assuming that the functional form of the metric remains of the Kerr type, where at each instant in time the mass and spin are governed by the BHSR process: in other words, that backreaction does not change the structure of the metric. This assumption is valid if the energy in each of the fields $\phi_i$ is sufficiently small, so that we can assume that the gravitational sector is still described by Einstein's equations in vacuum and the boson fields $\phi_i$ are propagating on a Kerr geometry. More rigorously, although the total mass of the scalar cloud can reach a significant fraction of the BH mass, the quantity which directly couples to the geometry through Einstein's equations is the energy density of the scalar cloud, which is suppressed by at least 8 orders of magnitude compared to the energy density associated to the BH event horizon $\rho_{\rm BH} \sim 1/(G^3M^2)$~\cite{Brito:2014wla,Stott:2018opm}. The reason is that the energy of the scalar cloud is spread over a huge volume, as the cloud itself peaks at a distance from the BH $r \sim 1/(G\mu^2M)$~\cite{Arvanitaki:2010sy,Brito:2014wla}. Therefore, the scalar cloud exerts a negligible gravitational pull on the background space-time geometry, so that backreaction effects can safely be neglected, as argued in a number of earlier works~\cite{Arvanitaki:2010sy,Brito:2014wla,Stott:2018opm}.

Although a full assessment of the effects of backreaction on the superradiant process is well beyond the scope of the present work, we wish to briefly comment on this issue. The boson cloud is physically extended over a scale $R_B \sim 1/(r_g\mu^2)$ which, using Eq.~(\ref{eq:alpha}), we can express as $R_B \sim r_g/\alpha^2$. Therefore, even for the largest value of $\alpha \approx 0.31$ considered here, the cloud is diluted over a region much larger than the BH gravitational radius $r_g$. We therefore expect the gravitational effect of the boson cloud on associated quantities, such as the BH shadow size, to be negligible. The only instance in which the cloud can have a significant effect on the shadow is if backreaction itself is non-negligible.

More generally, backreaction becomes important when the energy density of the boson cloud $\rho_B \sim M_B/R_B^3 \sim M_B\alpha^6/(GM)^3$ exceeds the energy density associated to the BH event horizon $\rho_{\rm BH} \sim 1/(G^3M^2)$. Therefore, backreaction is only important when $M_B \gg M/\alpha^6$, where $M_B$ is the total mass of the boson cloud, and $1/\alpha^6 \gtrsim 10^3$ for the regime we are considering. Satisfying the condition for which backreaction is important would require moving well beyond the region of parameter space we are considering, to a region where the modelling of the superradiance cannot be tackled analytically. In all the cases we examine, it is instead always the case that $M_B \lesssim M$. Therefore, we neglect backreaction in our study. In closing, we note other works such as Refs.~\cite{East:2013mfa,East:2017ovw} have studied superradiance in dynamical spacetimes, through a full numerical solution to the Einstein(-Proca) equations. We plan to return to this issue in future follow-up work.

\section{Methodology}
\label{sec:methodology}

To track the evolution of the BH shadow throughout the superradiant phase in the presence of gas accretion and GW emission, we solve the coupled system of differential equations for the BH mass and spin given by Eqs.~(\ref{eq:dotM},\ref{eq:dotJ}). In practice, however, we have found it convenient to re-scale the quantities of interest, so as to avoid dealing with extremely large or extremely small quantities, which could lead to numerical instabilities.

In particular, we write the boson occupation number as $N_i = N_{0i}\,n_i$, where $N_{0i} \equiv M_\odot/\mu_i$, and we define the mass of the BH as $M = M_\odot\,\tilde{M}$, so that $\tilde{M}$ is dimensionless and numerically equivalent to the value of the BH mass in units of Solar mass. Substituting the time derivative of the dimensionless spin $a_{\star}$ into Eq.~(\ref{eq:dotJ}) leads to the following set of coupled differential equations:
\begin{eqnarray}
\frac{\mathrm{d}n_i}{\mathrm{d}\tau} &=& \tilde{\Gamma}_in_i\,,
\label{eq:dotN2}\\
\frac{\mathrm{d}\tilde M}{\mathrm{d}\tau} &=& -\sum_i \left ( \tilde{\Gamma}_in_i + \tilde{\Gamma}_{{\rm GW},i}\,n_i^2 \right) + \tilde{\Gamma}_{\rm ACC} \,,
\label{eq:dotM2}\\
\frac{\mathrm{d}a_{\star}}{\mathrm{d}\tau} &=& - \frac{1}{\tilde M^2}\sum_i \frac{1}{\tilde\alpha_i}\, \left( j\tilde{\Gamma}_in_i + \tilde{\Gamma}_{{\rm GW},i}\,n_i^2 \right)\\
&&+ \frac{\tilde{\Gamma}_{\rm ACC}}{\tilde M}\!\left(\!\frac{\ell(t)}{E(t)} \!-\! 2a_\star\right) \!+\! \frac{2a_{\star}}{\tilde M}\sum_i \left ( \tilde{\Gamma}_i n_i \!+\! \tilde{\Gamma}_{{\rm GW},i}n_i^2 \right ) \,,\nonumber
\label{eq:dota2}
\end{eqnarray}
where we have defined $\ell \equiv L/M$ and $\tilde\alpha_i \equiv G\,M_\odot\,\mu_i$. The following quantities have also been introduced:
\begin{eqnarray}
\tilde \Gamma_{{\rm GW},i} &\equiv& \frac{P_{{\rm GW},i}}{P_{\rm Pl}\,n_i^2} = \frac{484+9\pi^2}{23040} \tilde\alpha_i^{14}\,\tilde M^{12}\,,
\label{eq:tildegammagwi}\\
\tilde \Gamma_{\rm ACC} &\equiv& \frac{2\times 10^{-8}\,{\rm yr}^{-1}}{\Gamma_{\rm Pl}}f_{\rm Edd}\,\tilde M\,,
\label{eq:tildegammaacc}\\
\tilde\Gamma_i &=& \frac{1}{24} \left [ a_{\star}-2\tilde\alpha_i\tilde{M}(1+\sqrt{1-a_{\star}^2}) \right ] \tilde\alpha_i^9\tilde M^8\,,
\label{eq:tildegammai}
\end{eqnarray}
where $\Gamma_{\rm Pl} \equiv M_{\rm Pl}/(M_\odot t_{\rm Pl}) \approx 6.402\times 10^{12}\,{\rm yr}^{-1}$ and $\tau = \Gamma_{\rm Pl}t$. We solve the set of differential equations given by Eqs.~(\ref{eq:dotN2}--\ref{eq:dota2}), with the following set of initial conditions: $M(t=0) = M_0$, $a_{\star}(t=0) = 0.99$, and $n_i(t=0) = 0$.

To plot the contour of the BH shadow, we use Eqs.~(\ref{eq:xietaxy}) in combination with a closed-form expression for the constants of motion $(\xi_c,\eta_c)$ characterizing unstable circular orbits, given by (see e.g.\ Ref.~\cite{Bambi:2008jg}):
\begin{eqnarray}
\xi_c(r,M,a) &=& \frac{M\,(r^2-a^2)-r(r^2-2Mr+a^2)}{a(r-M)}\,,
\label{eq:xic}\\
\eta_c(r,M,a) &=& \frac{4a^2Mr^3-r^4(r-3M)^2}{a^2(r-M)^2}\,.
\label{eq:etac}
\end{eqnarray}
Combining Eqs.~(\ref{eq:xietaxy},\ref{eq:xic},\ref{eq:etac}), we produce a parametric plot of $\{x(r),y(r)\}=\{x(\xi_c(r)),y(\xi_c(r),\eta_c(r))\}$ using Eq.~(\ref{eq:xietaxy}), with the parameter governing the plot being $r$, and domain chosen such that $y^2(r) \geq 0$. Note that throughout the superradiant phase, the BH shadow evolves due to the changes in the BH mass and spin.

We denote the total evolution timescale by $t_{\rm long}$, and define it to be the time taken by the BH to evolve from its initial state with dimensionless spin $a_{\star}=0.99$, to the minimum dimensionless spin allowed by the BHSR process $a_{\star}=a_{\min}^{(1)}$.\footnote{While above we have provided a nominal definition of $t_{\rm long}$, for numerical computations we have found it more convenient to adopt an operational definition of $t_{\rm long}$ as the time it takes for the BH spin to evolve from its initial value to $100.05\%$ of the minimum spin $a^{(1)}_{\min}$. This is an operationally more efficient definition as the minimum spin is approached asymptotically, and integrating the BHSR equations up to $a^{(1)}_{\min}$ will result in a substantial amount of numerical uncertainty on the determined $t_{\rm long}$ as machine precision is reached. We have verified that this choice of spin cut-off in numerically computing $t_{\rm long}$ is meaningful throughout the parameter space for the relative mass parameter $\beta$ we explore. In other words, even in the most extreme regions of parameter space, such a definition always cuts off the BHSR evolution at a point where the spin evolution has already plateaued significantly: this might not be the case for less conservative cut-off choices, which in some cases may lead to a shorter $t_{\rm long}$, at the cost of the BHSR process not having slowed down sufficiently when the spin cut-off occurs.} Since $a_{\min}^{(i)}$ is monotonically increasing with $\alpha_i$, the gravitational coupling of the lighter species determines the final spin of the BH. The change in the shadow of the BH, on the other hand, is characterized by $\Delta R$ and $\Delta \chi$: these are defined as the differences in the quantities $R$ and $\chi$ [given in Eqs.~(\ref{eq:diameter},\ref{eq:chi})] between $t=t_{\rm long}$ and $t=0$. In practice, $\Delta R$ is not a directly observable quantity, as it characterizes a change in the size of the BH shadow in units of BH mass and does not take into consideration the distance to the observer. For this reason, we also define the observable quantity:
\begin{equation}
\label{eq:delRAbs}
\Delta R_{\rm Abs} \simeq 42{\rm\,\mu as}\,\bigg(\frac{\Delta R}{6\sqrt{3}}\bigg)\bigg(\frac{M_0}{6.5\times10^9\,M_\odot}\bigg)\bigg(\frac{16.8{\rm\,Mpc}}{D}\bigg)\,,
\end{equation}
which denotes the absolute change in the angular size of the shadow of a BH of initial mass $M_0$, as seen by an observer located at distance $D$ from the BH, reported in $\mu{\rm as}$. In general, the quantities $M_0$ and $D$ characterize the superradiance-induced BH shadow evolution along with $t_{\rm long}$. Note in addition that for numerical convenience we have rescaled Eq.~(\ref{eq:delRAbs}) using the values of $M_0$ and $D$ relevant for the SMBH M87$^*$.

\section{Results}
\label{sec:results}

Let us consider two boson species, with gravitational couplings $\alpha_1$ and $\alpha_2$, where $\alpha_2 \geq \alpha_1$. These gravitational couplings are defined with respect to the BH's initial mass (it is important to specify this point given that the BH's mass will vary in time due to BHSR). We define the relative difference between the boson masses through the relative mass parameter $\beta \equiv 1-\alpha_1/\alpha_2$. Throughout our analysis, we fix $\alpha_2=0.31$ and vary $\beta$ in order to scan across the parameter space for the mass of the lightest boson. Therefore, the limit $\beta \to 0$ effectively reproduces a single-species case with $\alpha=0.31$.~\footnote{Note that we fix the numerical value of $\alpha_2$ as defined in Eq.~(\ref{eq:alpha}), where the product $M\mu$ appears. Therefore, the mass of the boson changes when scanning over the BH mass range. This procedure is different from fixing the mass of the boson or, equivalently, the quantity $\tilde{\alpha}_2$ which appears in Eqs.~(\ref{eq:dotN2}--\ref{eq:dota2}).} The numerical value for the gravitational coupling $\alpha=0.31$ has been chosen in order to have a timescale $t_{\rm long}$ larger than $\sim 100\,$yrs and is based on the study in Ref.~\cite{Roy:2019esk}, where it is shown that for $\alpha = 0.32$ the timescale is $\sim 121\,$yrs, while for $\alpha = 0.28$ the timescale is $\sim 216\,$yrs.

Let us comment on our choices of masses/gravitational couplings of the two light boson species. Firstly, the BHSR equations we are solving are only valid in the linear regime $\alpha_1, \alpha_2 \lesssim {\cal O}(1)$. At the same time, the BHSR energy extraction process is most efficient for higher values of the gravitational coupling, see Eq.~(\ref{eq:gammai}). Therefore, by setting $\alpha_2=0.31$, we are placing ourselves in a ``sweet spot'' roughly at the edge of the validity for the linear regime while at the same time being the most optimistic from the observational point of view, given that the energy extraction is as efficient as possible.

We numerically solve the set of differential equations determined by Eqs.~(\ref{eq:dotN2}--\ref{eq:dota2}), with initial conditions given by $a_{\rm initial}=0.99$, $n_i=0$, and $\tilde{M}=M_0/M_\odot$ where the initial mass of the BH is $M_0$. We then use the functions $M(t)$ and $a_{\star}(t)$ to generate the contour of the BH shadow as a function of time, using Eqs.~(\ref{eq:xic},\ref{eq:etac}) with $a_{\star}=J/M^2$, as described in Sec.~\ref{sec:methodology}. From the generated shadows we extract the relevant observables $R$ and $\chi$ defined in Eqs.~(\ref{eq:diameter},\ref{eq:chi}), whose change in time we then analyze.

\subsection{Single-species case}
\label{subsec:note}

Before discussing the two-species case, let us point out an important subtlety pertaining to the single-species case, which we recall is reproduced by taking the limit $\beta \rightarrow 0$ (for a single boson species with gravitational coupling $\alpha=0.31$). The single-species case has been extensively studied in previous literature, including by one of us in Ref.~\cite{Roy:2019esk} and by Creci \textit{et al.}\ in Ref.~\cite{Creci:2020mfg}. Focusing on SgrA$^*$, the SMBH at the center of our galaxy, in Ref.~\cite{Roy:2019esk} it was shown that there exists a range of gravitational couplings where the BH shadow size will increase by ${\cal O}(0.1)\,\mu{\rm as}$ over a timescale potentially as short as ${\cal O}(100)\,{\rm yrs}$, which however is still longer than typical human timescales. Later, Ref.~\cite{Creci:2020mfg} analyzed a wider range of gravitational coupling parameter space, and showed that the BH shadow size can not only increase, but decrease as well, albeit still along the course of unrealistically long timescales.

While the most optimistic timescales reported in Ref.~\cite{Roy:2019esk} range around ${\cal O}(100)\,{\rm yrs}$, here we point out that there is a physically/observationally more relevant definition of the evolution timescale other than $t_{\rm long}$, which at the same time leads to more optimistic results. The key observation in this sense is that, while the \textit{complete} BHSR evolution does indeed take place over the timescale given by $t_{\rm long}$, most of the evolution actually takes place over a much shorter timescale, which we refer to as $t_{\rm short}$. For example, once the BH spin is sufficiently close to the final spin $a_{\min}$, the spin evolution slows down considerably and reaches a quasi-steady state, where the gradient ${\rm d}a_{\star}/{\rm d}t$ is very low compared to its value throughout the rest of the BHSR process. These considerations motivate us to define $t_{\rm short}$ as the time it takes for the BH spin to evolve from $99$\% of its initial spin $a_{\rm initial}$ to $101$\% of its final spin $a_{\min}$. This denotes the most significant period of the BHSR evolution process (over which $\approx 98\%$ of the spin evolution takes place), during which the gradient of the spin evolution against time is largest in magnitude, before later plateauing. Of course, by extension this is also the period over which the BH shadow size evolves the most over the shortest possible period of time.

\begin{figure*}[ht]
\centering
\begin{subfigure}{.48\textwidth}
\includegraphics[width=\textwidth]{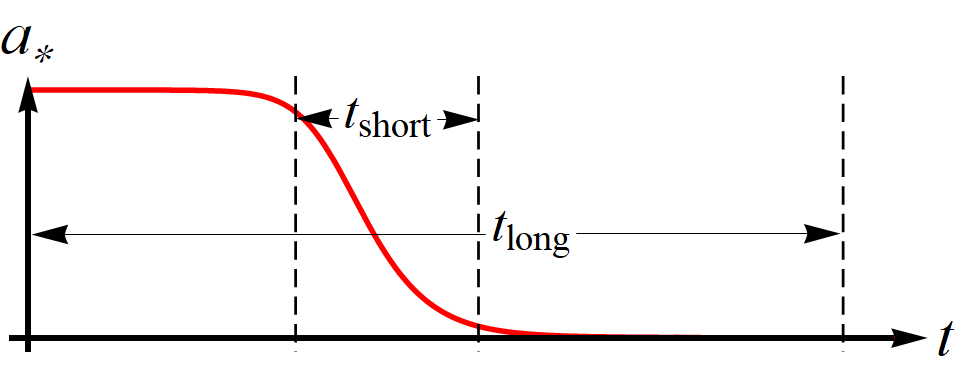}
\caption{Evolution of the dimensionless spin parameter as a function of time (in arbitrary units) in the presence of a single light boson species. We clearly highlight the difference between the total evolution timescale $t_{\rm long}$, and the shorter but observationally more relevant timescale $t_{\rm short}$. For the specific case of SgrA$^*$, the latter can be up to a factor of 5 shorter than the former, and comparable to human timescales.}
\label{fig:tshort1}
\end{subfigure}
\hfill
\begin{subfigure}{.48\textwidth}
\includegraphics[width=\textwidth]{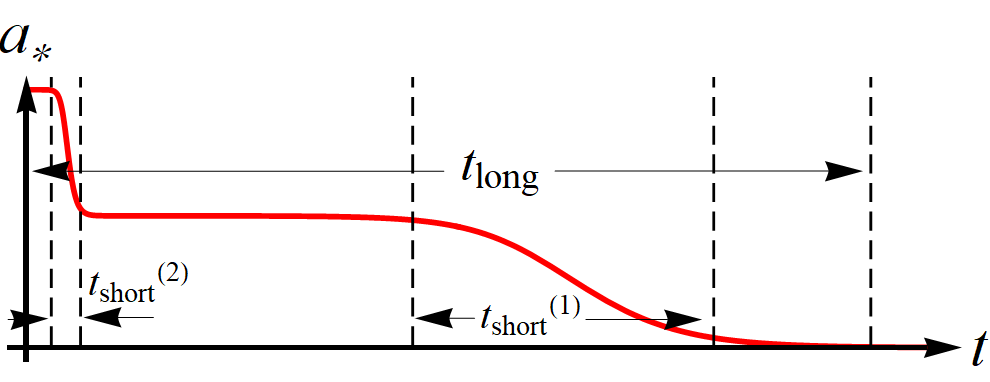}
\caption{Same as in panel (a) in the presence of two light boson species. The gravitational coupling of the heaviest one is $\alpha=0.31$ and the relative mass parameter is $\beta=0.2$, such that the contribution of both species to the BH spin extraction is comparable. Here, one can define two short timescales $t_{\rm short}^{(2)}$ and $t_{\rm short}^{(1)}$, associated to the heaviest and lightest species respectively.}
\label{fig:tshort2}
\end{subfigure}
\hfill
\begin{subfigure}{.48\textwidth}
\includegraphics[width=\textwidth]{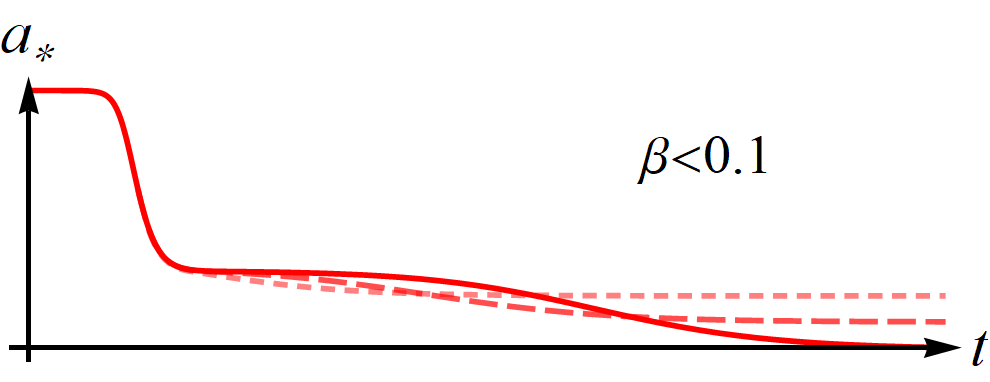}
\caption{Same as in panel (b), with relative mass parameters given by $\beta=0.03$ (dotted), $\beta=0.06$ (dashed), and $\beta=0.09$ (solid). Within this regime, the dynamics of the system are mostly driven by the heavier of the two species, as $p_1 \ll p_2$ with the $p_i$s defined in Eqs.~(\ref{eq:pi}). Therefore, the significant evolution timescale is $t_{\rm short}^{(2)}$.}
\label{fig:tshort3}
\end{subfigure}
\hfill
\begin{subfigure}{.48\textwidth}
\includegraphics[width=\textwidth]{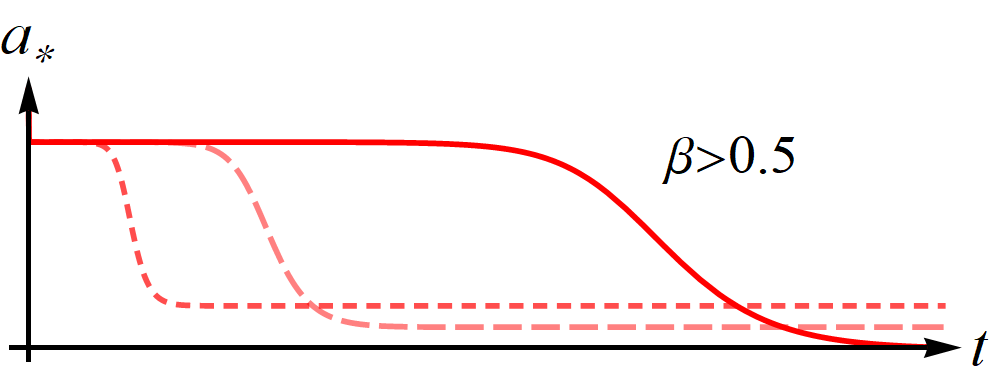}
\caption{Same as in panel (b), with relative mass parameters given by $\beta=0.5$ (dotted), $\beta=0.55$ (dashed) and $\beta=0.6$ (solid). Within this regime, the dynamics of the system are mostly driven by the lightest of the two species, as $p_2 \ll p_1$ with the $p_i$s defined in Eqs.~(\ref{eq:pi}). Therefore, the significant evolution timescale is $t_{\rm short}^{(1)}$.}
\label{fig:tshort4}
\end{subfigure}
\caption{The different panels show the evolution of the dimensionless spin parameter as a function of time (in arbitrary units) for various single- and two-species light boson systems, highlighting the difference between $t_{\rm long}$ and $t_{\rm short}$, and the fact that within certain regimes for the relative mass parameter $\beta$, the dynamics of the system are mostly driven by one of the two species, either the heavier one ($\beta<0.1$) or the lighter one ($\beta>0.5$). See the captions of the specific panels for more details.}
\label{fig:spina}
\end{figure*}

The various sub-panels of Fig.~\ref{fig:spina} show the evolution of the dimensionless spin against time, in arbitrary units, for various values of $\beta$.~\footnote{We stress that our choice of removing numbers from the axes, and therefore considering arbitrary units, is deliberate. The aim of Fig.~\ref{fig:spina} is purely illustrative, the goal being that of providing the reader with a visual representation of the qualitative behaviour of the system for different regimes of $\beta$, which we will discuss at a quantitative level in more detail later in the text.} More specifically, in Fig.~\ref{fig:tshort1} we show this evolution in the presence of a single boson species, showcasing the difference between $t_{\rm short}$ and $t_{\rm long}$. From Fig.~\ref{fig:tshort1}, it is clear that $t_{\rm short}$ is the observationally more relevant timescale, as it is the period throughout which an instrument has the best chance of observing significant changes (over a short period of time) in the BH shadow. The cutoff that defines $t_{\rm short}$ occurs right before the quasi-steady state of the BH spin evolution begins. Note that $t_{\rm short}$ is shorter than the period defined as ``Phase I'' in Ref.~\cite{Creci:2020mfg}. Figs.~\ref{fig:tshort2},~\ref{fig:tshort3}, and~\ref{fig:tshort4} are instead devoted to the two-species case, and will be discussed in more detail in Sec.~\ref{subsec:multiplespecies}.

\subsection{Two-species case}
\label{subsec:multiplespecies}

\begin{figure*}
\begin{tabular}{cc}
\begin{subfigure}{.4\textwidth}
\includegraphics[width=\textwidth]{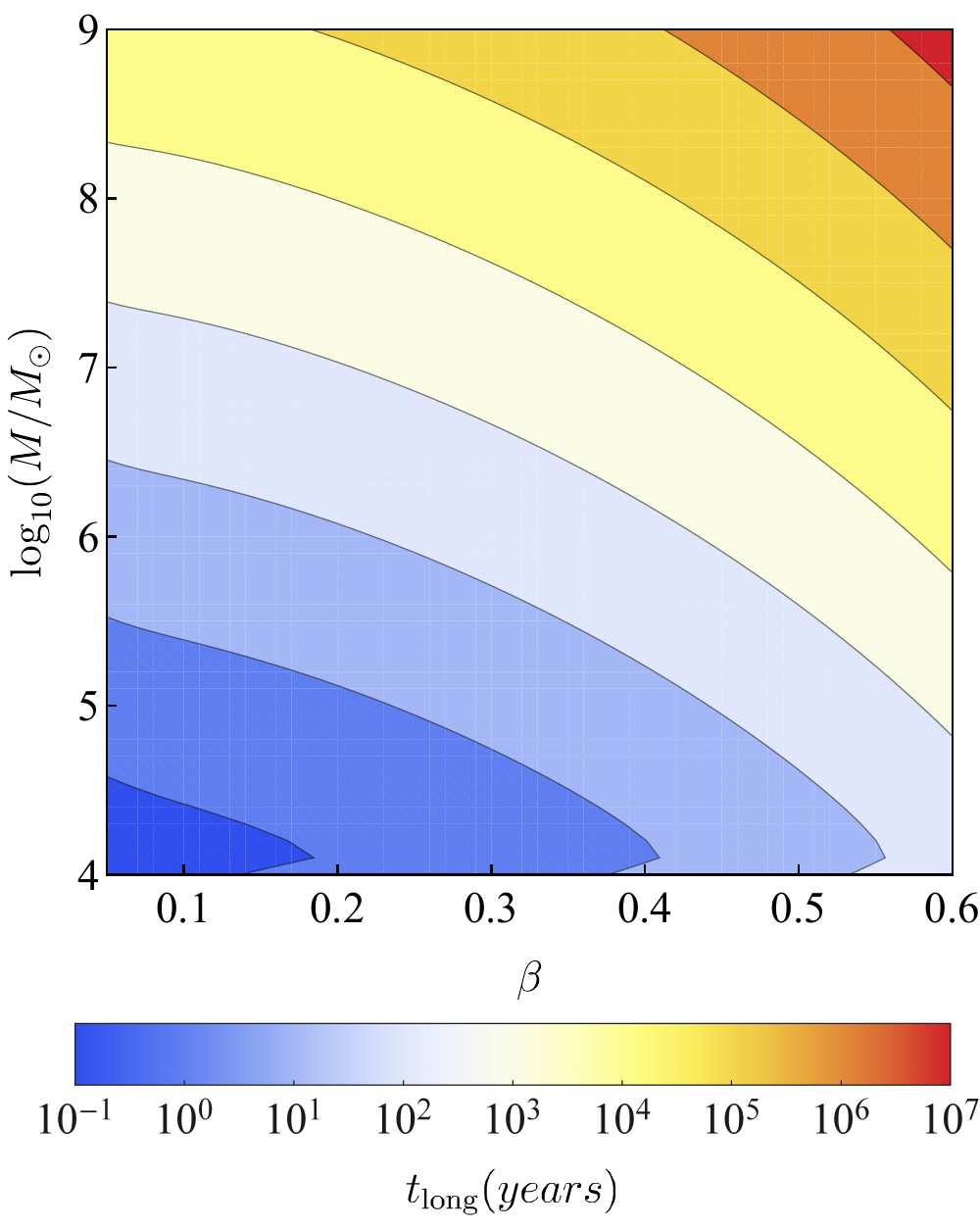}
\caption{Total evolution timescale $t_{\rm long}$ as a function of $\log_{10}(M/M_{\odot})$ and $\beta$. This is the time taken by the BH's dimensionless spin to evolve from $a_{\star}=0.99$ to the minimum value allowed by the BHSR process.}
\end{subfigure} &\quad \quad \quad
\begin{subfigure}{.4\textwidth}
\includegraphics[width=\textwidth]{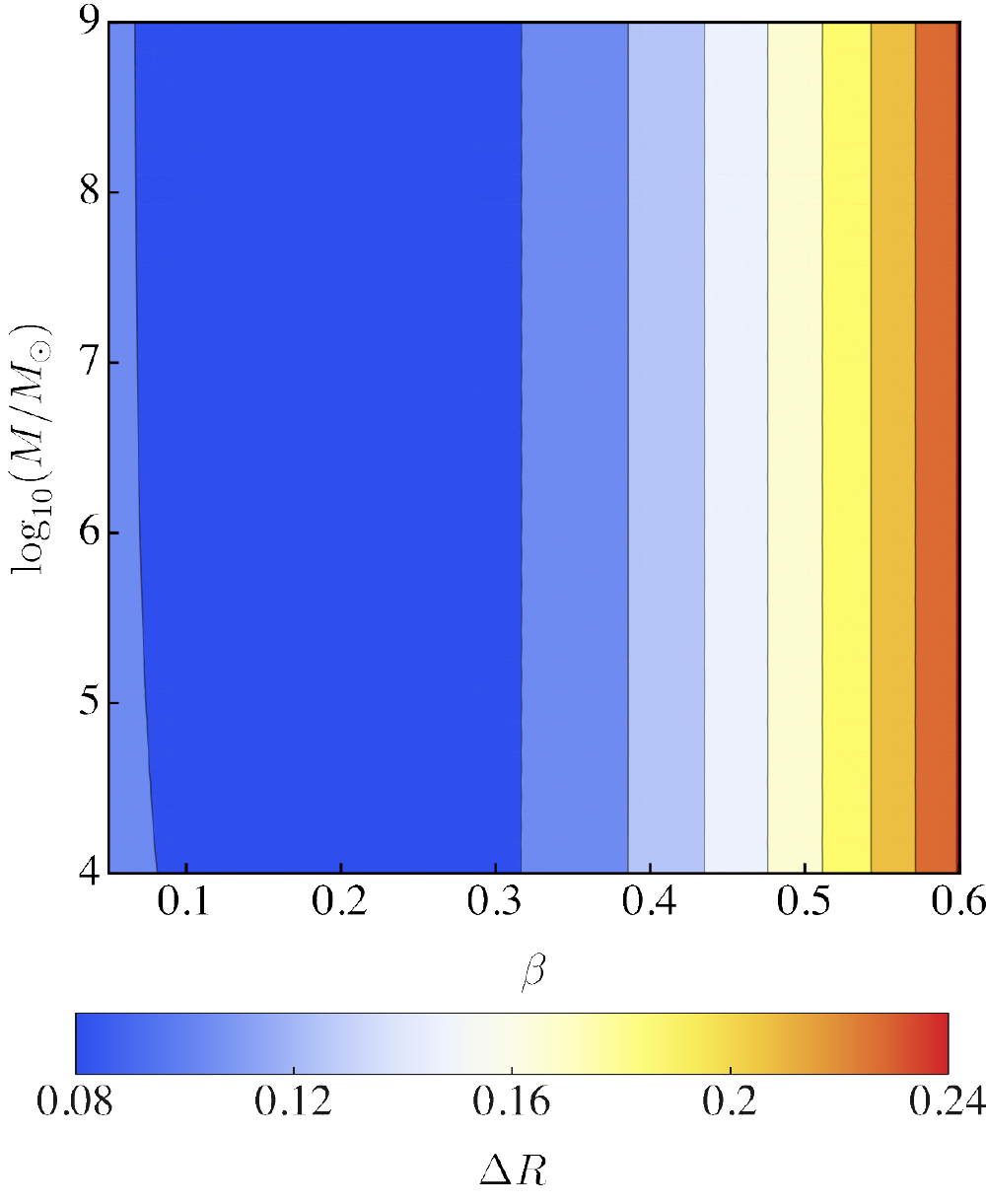}
\caption{Net change in the diameter of the shadow $\Delta R$ as the system evolves from $t=0$ to $t=t_{\rm long}$, in units of the initial BH mass $M$, as a function of $\log_{10}(M/M_{\odot})$ and $\beta$.}
\end{subfigure}\\
\begin{subfigure}{.4\textwidth}
\includegraphics[width=\textwidth]{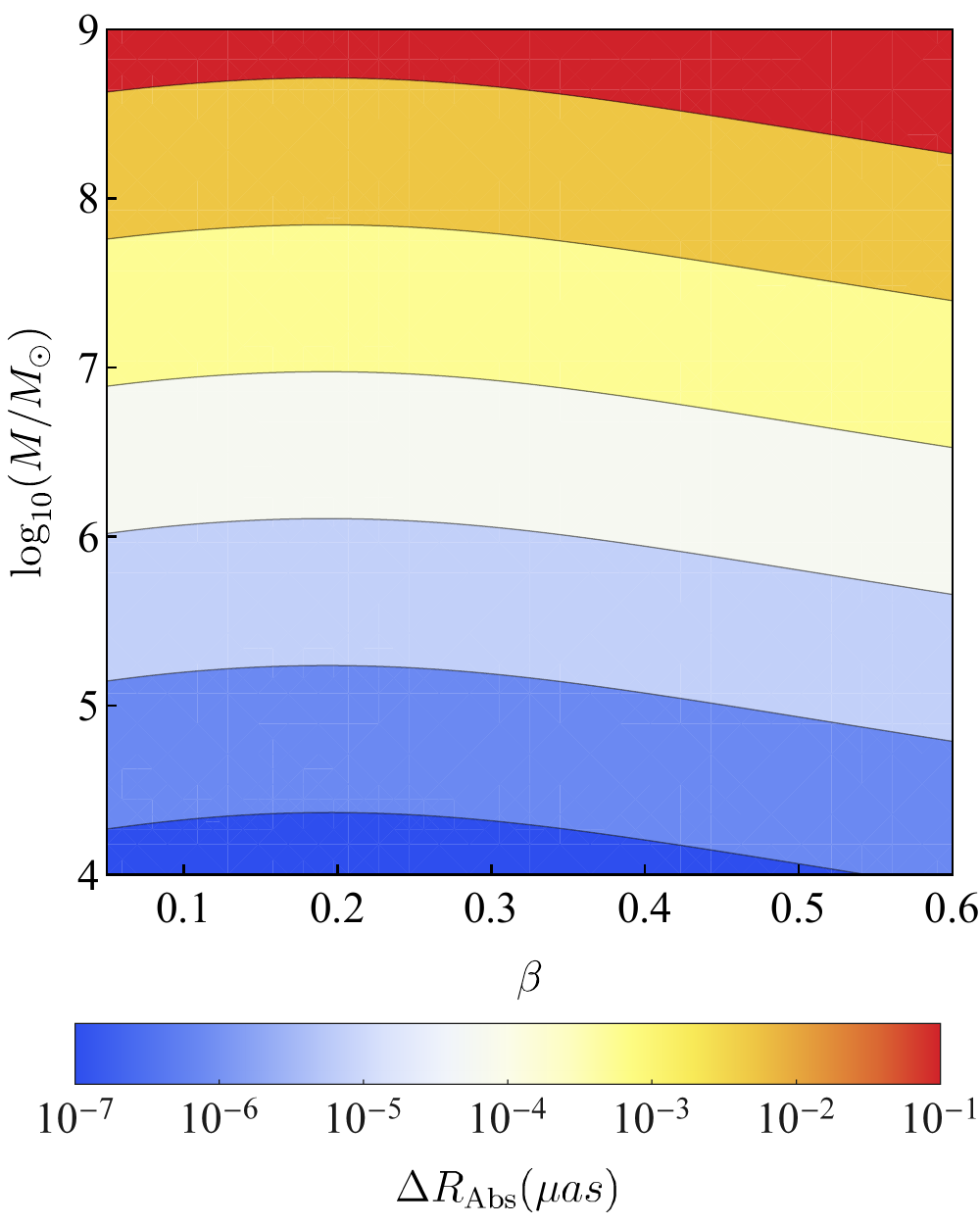}
\caption{Net change in the angular size of the shadow $\Delta R_{\rm Abs}$ as the system evolves from $t=0$ to $t=t_{\rm long}$ seen by an observer at a distance $D=16.8\,{\rm Mpc}$, as a function of $\log_{10}(M/M_{\odot})$ and $\beta$.}
\end{subfigure} & \quad \quad \quad
\begin{subfigure}{.4\textwidth}
\includegraphics[width=\textwidth]{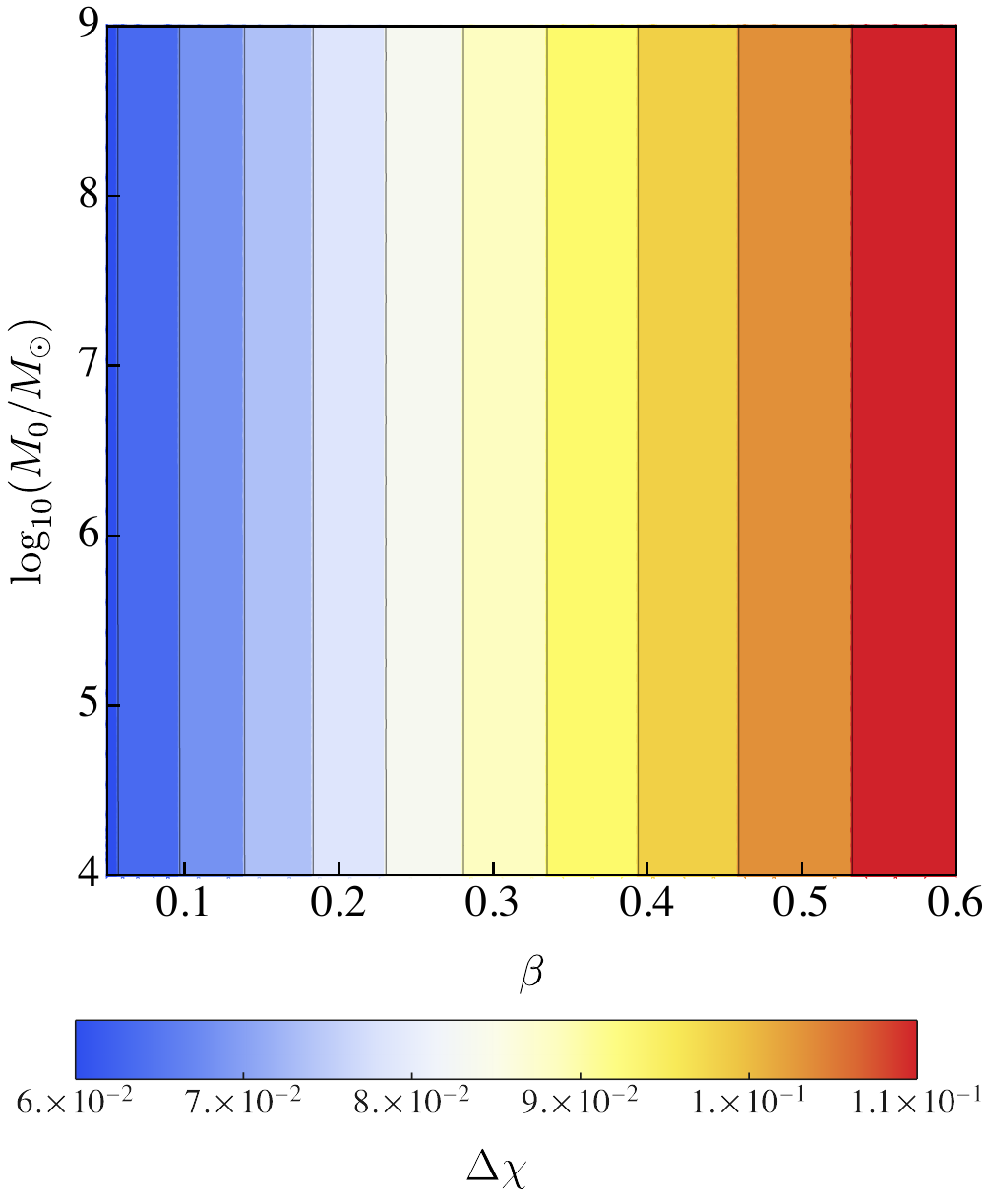}
\caption{Net change in the axis ratio of the shadow $\Delta \chi$ (which quantifies the deviation from circularity of the shadow) as the system evolves from $t=0$ to $t=t_{\rm long}$, as a function of $\log_{10}(M/M_{\odot})$ and $\beta$.}
\end{subfigure}
\end{tabular}
\caption{Contour plots for various observables relevant to the superradiance-induced BH shadow evolution as a function of the logarithm of the initial mass of the BH, $\log_{10}(M/M_{\odot})$ (horizontal axis) and the relative mass parameter $\beta \equiv 1-\alpha_2/\alpha_1$ (vertical axis). The quantities $\alpha_2=0.31$ and $\alpha_1$ are the gravitational couplings of the heaviest and lightest boson species respectively. For each sub-panel, we show contours of fixed logarithmically equispaced values for the observable in question. See the captions of the specific panels for more details.}
\label{fig:densityPlot}
\end{figure*}

Let us now consider two boson species with gravitational couplings $\alpha_1$ and $\alpha_2$, and relative mass parameter $\beta \equiv 1-\alpha_1/\alpha_2$. Fig.~\ref{fig:densityPlot} provides contour plots of various relevant observational quantities (total evolution timescale $t_{\rm long}$, change in BH shadow angular size in units of BH mass $\Delta R$, absolute change in BH shadow angular size in $\mu{\rm as}$ considering a BH located at a distance $D=16.8\,{\rm Mpc}$ from us as is the case for M87$^*$, and change in axis ratio $\Delta \chi$) across a grid of the (logarithm of the) initial BH mass $\log_{10}M_0$ and the relative mass parameter $\beta$. We set the upper limit $\beta<0.6$ in order to ensure that \textit{a)} the total evolution timescale $t_{\rm long}$ is not too large (to the point of being unobservable), and \textit{b)} the superradiance rate for both boson species dominates over the gas accretion and GW emission rates.

Ideally, for the BHSR-induced changes in the BH shadow to be as observationally favorable as possible, we would like the associated changes in the shadow parameters ($\Delta R$, $\Delta R_{\rm Abs}$, and $\Delta \chi$) to be as large as possible, with $t_{\rm long}$ being as short as possible. As is clear from the top left panel of Fig.~\ref{fig:densityPlot}, $t_{\rm long}$ increases when increasing both the initial mass of the BH $M_0$ and the relative mass $\beta$. At first glance, the behavior of $t_{\rm long}$ as a function of $\beta$ may seem counterintuitive, as one might expect a system with multiple boson species to evolve more quickly than one with a single species: in fact, besides theoretical (beyond the SM) considerations, this was one of our original motivations for considering a multiple-species boson system, expecting the resulting BHSR-induced evolution to possibly be quicker, and hence observationally more favorable. However, the physical reason for this counterintuitive result can be understood by looking at the superradiance rate as given by Eq.~(\ref{eq:gammai}). Since $\Gamma_i$ depends on the spin parameter $a_{\star}$, when any one of the two boson species starts extracting angular momentum from the BH, the superradiance rate for both species decreases and hence makes the energy extraction process overall less efficient. This counterbalances the possible quicker evolution, with the final result being that the two-species evolution ends up proceeding over a longer timescale compared to the single-species case.

So far, we have only discussed the evolution of $t_{\rm long}$ with $\beta$. We now discuss how the absolute change in the BH shadow size $\Delta R_{\rm Abs}$ and the change in the axis ratio $\Delta \chi$ respond to $\beta$. We notice that $\Delta R_{\rm Abs}$ initially decreases as $\beta$ increases, reaching a minimum at $\beta \approx 0.2$ before increasing again with increasing $\beta$. As we will discuss in more detail later in this Section, this behavior can be explained in terms of the competing effects on the BH shadow size due to the decrease in BH spin and mass. Of course, $\Delta R_{\rm Abs}$ increases with increasing $M_0$, since the BH shadow size is directly proportional to the BH mass: this behavior is clearly visible in the lower left panel of Fig.~\ref{fig:densityPlot}. It might therefore be desirable to factor out this extra dependence on the BH initial mass, by considering the change in shadow size in units of BH mass $\Delta R$: the result, clearly independent of $M_0$, is shown in the top right panel of Fig.~\ref{fig:densityPlot}. The change in the the BH shadow axis ratio $\Delta \chi$ is to very good approximation independent of the initial BH mass, but is found to increase as $\beta$ increases. However, for all values of $\beta$ we explored, $\Delta\chi$ remains small. $\Delta\chi \lesssim {\cal O}(0.1)$. This figure is about one order of magnitude worse than current limits on $\Delta \chi \lesssim 4/3$,  and also likely beyond the reach of near-future VLBI arrays (although to the best of our knowledge projections in this sense are not available).

\begin{figure}
\includegraphics[width=1.0\linewidth]{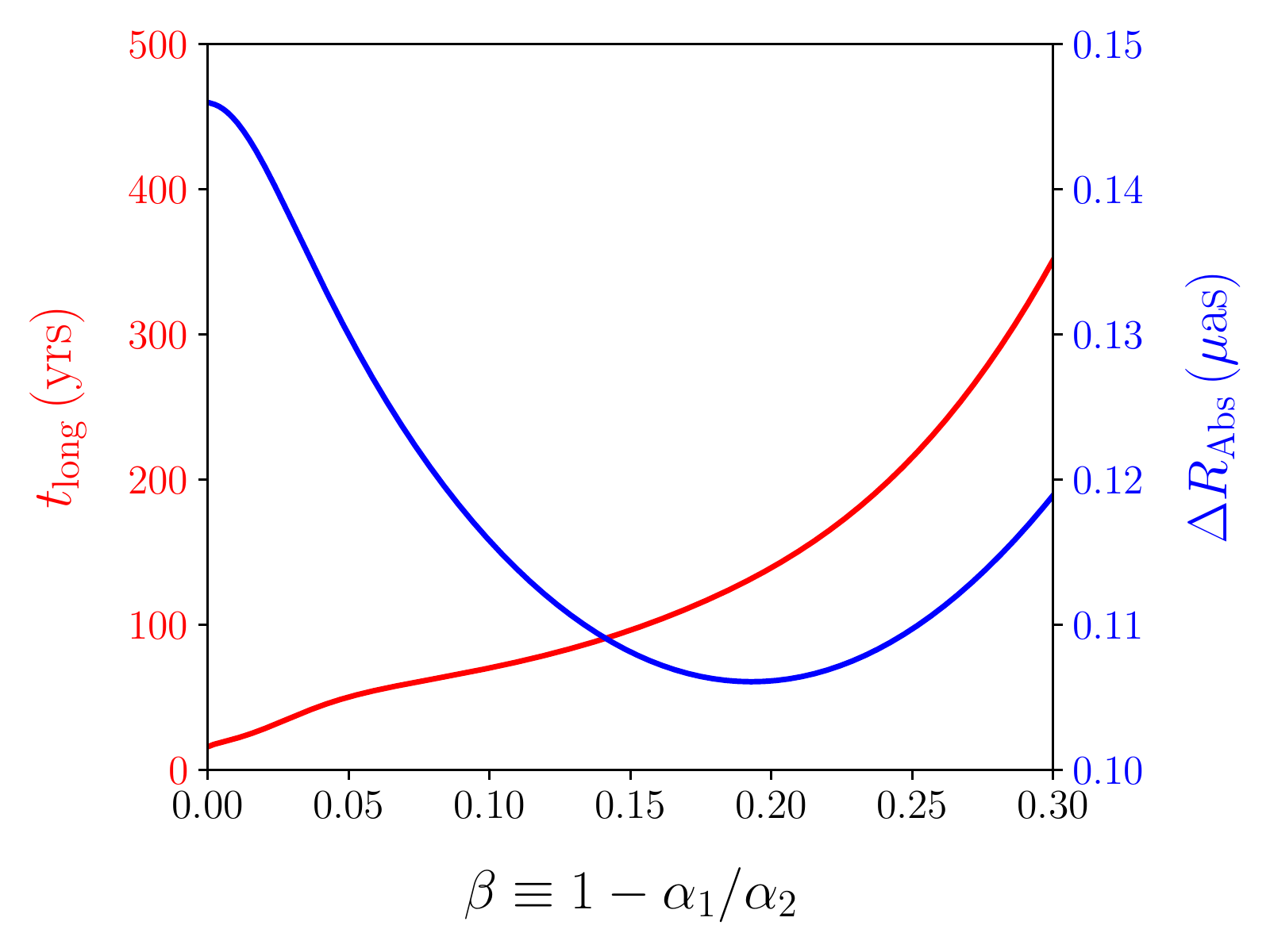}
\caption{Total evolution timescale $t_{\rm long}$ (red, left vertical axis, in years) and net change in the angular size of the BH shadow $\Delta R_{\rm Abs}$ (blue, right vertical axis, in $\mu{\rm as}$) as a function of the relative mass parameter $\beta$. In producing this plot, the BH parameters have been fixed to those of SgrA$^*$. The plot clearly shows the trade-off between the parameters $\Delta R_{\rm abs}$ and $t_{\rm long}$, given that observationally favorable conditions would want the former to be as large as possible with the latter being as small as possible.}
\label{fig:tshort_DeltaRabs_vs_beta_log}
\end{figure}

The previous results indicate that, from the point of view of BH shadow-related observational quantities ($\Delta R_{\rm Abs}$ or $\Delta R$, and $\Delta \chi$), the best prospects for detecting the effects of BHSR in the two-species case would appear to point towards higher values of $\beta \gtrsim 0.2$. However, as we saw earlier, the evolution timescale increases monotonically with $\beta$. Focusing on $\beta$ as independent parameter, there is therefore a trade-off between evolution timescale and the previously discussed quantities: we show this in Fig.~\ref{fig:tshort_DeltaRabs_vs_beta_log}, where we plot the evolution of both $t_{\rm long}$ and $\Delta R_{\rm Abs}$ against $\beta$ for the specific case of SgrA$^*$, clearly displaying the behaviours discussed previously.

In Sec.~\ref{subsec:note}, we argued that the single-species is more subtle than reported in previous works, and requires a careful definition of the observationally relevant timescale. What about the two-species case? As for the single-species case, we can in principle define a short significant evolution timescale for each species, $t_{\rm short}^{(i)}$. Operationally, $t_{\rm short}^{(2)}$ and $t_{\rm short}^{(1)}$ are defined as the times taken for the BH spin to evolve from $99\%$ of $a_{\rm initial}$ to $101\%$ of $a_{\min}^{(2)}$ and from $99\%$ of $a_{\min}^{(2)}$ to $101\%$ of $a_{\min}^{(1)}$ respectively (see Fig.~\ref{fig:tshort2}, in arbitrary units). The evolution of the BH shadow is of course distributed over the entire evolution timescale, and hence in principle we expect both $t_{\rm short}^{(2)}$ and $t_{\rm short}^{(1)}$ to be observationally significant.

However, in the case of small ($\beta \lesssim 0.05$) and large ($\beta \gtrsim 0.5$) relative mass, it turns out that only one of the two short timescales $t_{\rm short}^{(i)}$ is observationally relevant. To better quantify this, we define the contribution fraction $p_i$ ($i=1,2$), which quantifies the fraction of the BH spin evolution contributed by each species throughout the BHSR process. We define the $p_i$s as:
\begin{eqnarray}
p_1 = \frac{a_{\min}^{(2)}-a_{\min}^{(1)}}{a_{\rm initial}-a_{\min}^{(1)}}\,, \quad \quad
p_2 = \frac{a_{\rm initial}-a_{\min}^{(2)}}{a_{\rm initial}-a_{\min}^{(1)}}\,\label{eq:pi}.
\end{eqnarray}
When $\beta \lesssim 0.05$, the mass parameters of the two species are very close to each other, and hence $a_{\min}^{(2)}-a_{\min}^{(1)} \ll a_{\rm initial}-a_{\min}^{(1)}$. This implies $p_1 \ll p_2$, and therefore the significant evolution timescale is mostly driven by the heavier species: see Fig.~\ref{fig:tshort3}, in arbitrary units. The converse occurs for $\beta \gtrsim 0.5$: in this case, $a_{\min}^{(2)}-a_{\min}^{(1)} \gg a_{\rm initial}-a_{\min}^{(2)}$, implying $p_2 \ll p_1$, and therefore it is the lightest species which is mostly driving the significant evolution timescale: see Fig.~\ref{fig:tshort4}, again in arbitrary units.

It is worth noting that, in each of these limiting cases (both large and small $\beta$), the behavior of the system is effectively very close to that of the single-species case, although the underlying physical reason is different in the two regimes. In the small $\beta$ limit, the behavior is very similar to that of a single-species system precisely because the masses of the two species are very close to each other. In the large $\beta$ limit, this is instead the case because the lighter species dominates over the heavier one as far as the BHSR process is concerned. On the other hand in the intermediate region, $0.05 \lesssim \beta \lesssim 0.5$, the contribution from both species is comparable.

By combining Eqs.~(\ref{eq:amin},~\ref{eq:pi}) for $\beta=0.2$, the two contribution fractions are equal to each other: $p_1=p_2=0.5$. From the point of view of energy extraction, this would therefore na\"{i}vely appear to be the observationally most optimistic case. However, as noted earlier and shown in Fig.~\ref{fig:tshort_DeltaRabs_vs_beta_log}, $\Delta R_{\rm Abs}$ actually dips and reaches a minimum for $\beta \approx 0.2$. A physical explanation for this phenomenon is as follows: as far as the absolute size of the BH shadow is concerned, the BHSR-induced mass and spin extraction lead to competing effects. Decreasing the spin makes the BH shadow more circular, and hence increases its size (at fixed BH mass). On the other hand, decreasing the mass scales down the entire BH shadow size, as the latter is (neglecting spin effects) proportional to the BH mass throughout the BHSR process. For $\beta \approx 0.2$, the contribution of both species is similar, and therefore the competing effects of mass and spin extraction is the largest. This leads to an overall smaller change in the BH shadow size in comparison to the changes obtained for both smaller and larger values of $\beta$.

\section{Discussion}
\label{sec:discussion}

Our previous discussion of the BHSR-induced evolution of the BH shadow (in both the single- and two-species case) and the associated evolution timescales has been rather general. To further assess the observational prospects of the effects discussed, let us now consider a specific worked example focused on SgrA$^*$, the SMBH at the center of our galaxy, with a mass of $M=4.2 \times 10^6\,M_\odot$ and located at a distance of $D=8\,{\rm kpc}$ from us~\cite{2019A&A...625L..10G}. For small values of $\beta<0.05$, we present the results of this analysis in Table~\ref{tab:SgrA}.

\begin{table}[!t]
\def\arraystretch{1.5}
\setlength{\tabcolsep}{12pt}
\begin{tabular}{cccc}
\hline\hline
$\beta$ & $t_{\rm long}$\,(yrs) & $\Delta \chi$ & $\Delta R_{\rm Abs}$\,($\mu{\rm as}$)\\
\hline
$10^{-3}$ & 73  & 0.06 & 0.62\\
0.01 & 97  & 0.06 & 0.61\\
0.03 & 172  & 0.06 & 0.57\\ 
0.05 & 233  & 0.06 & 0.54\\ 
\hline\hline
\end{tabular}
\caption{Total evolution timescale $t_{\rm long}$, net change in the axis ratio $\Delta \chi$, and net change in the angular size of the BH shadow $\Delta R_{\rm Abs}$ as a function of the relative mass parameter $\beta$. These values have been computed for the specific case of SgrA$^*$, located at a distance $D=8\,{\rm kpc}$ away from us and with a mass of $M=4.2 \times 10^6 M_\odot$. While the total evolution timescale for all the cases reported exceeds typical human timescales, the shorter but observationally more relevant timescale $t_{\rm short}$ is $\sim 16\,{\rm yrs}$ for all cases reported in the Table.}
\label{tab:SgrA}
\end{table}

For the single-species case, the total evolution timescale $t_{\rm long}$ is reported in the first row of Table~\ref{tab:SgrA}: in this case we set the relative mass to $\beta=10^{-3}$, sufficiently small for this to be effectively considered a single-species case. We find the total evolution timescale to be $t_{\rm long} \simeq 70\,{\rm yrs}$. On the other hand, we find the observationally significant timescale $t_{\rm short}$ to be a factor of $\sim 5$ shorter, $t_{\rm short} \simeq 16 \,{\rm yrs}$. Throughout the timescale characterized by $t_{\rm short}$, the absolute angular size of the shadow changes by $\Delta R_{\rm Abs} \sim 0.6\,\mu{\rm as}$. This figure is still about a factor of $\sim 5$ below the resolution of planned near-future space-based VLBI arrays~\cite{Fish:2019epg}.

As $\beta$ is increased (see subsequent rows in Table~\ref{tab:SgrA}), $t_{\rm long}$ correspondingly increases and the absolute change in the shadow angular size $\Delta R_{\rm Abs}$ decreases, confirming the behavior we saw earlier (see Fig.~\ref{fig:tshort_DeltaRabs_vs_beta_log}). However, as discussed in Sec.~\ref{subsec:multiplespecies} (see also Fig.~\ref{fig:tshort3}), for sufficiently small $\beta$ the behavior of the system is still akin to the single-species case, with dynamics driven by the heavier species. Therefore, in spite of the significant increase in $t_{\rm long}$ as $\beta$ is increased, is might be instructive to focus on the short timescale associated to the heavier species, $t_{\rm short}^{(2)}$, rather than on the total evolution timescale $t_{\rm long}$. We find that, for all the values of $\beta$ reported in the Table, $t_{\rm short}^{(2)} \simeq 16\,{\rm yrs}$, well within typical human timescales.

Once we increase $\beta$ within the range $0.05 \lesssim \beta \lesssim 0.5$, where the contribution to the BH shadow evolution from both species is comparable as discussed earlier in Sec.~\ref{sec:discussion}, the distinction between the two short evolution timescales $t_{\rm short}^{(1)}$ and $t_{\rm short}^{(2)}$ becomes significantly murkier: at this point, there is a substantial degree of overlap between the two species since neither of them is clearly providing a dominant contribution to the evolution. More concretely, $p_1$ and $p_2$ as defined in Eq.~(\ref{eq:pi}) are roughly of the same order. Therefore, we conclude that for larger values of $\beta$ the more relevant timescale is actually the complete evolution timescale $t_{\rm long}$.

At the same time, as is clear from Table~\ref{tab:SgrA} and Fig.~\ref{fig:tshort_DeltaRabs_vs_beta_log}, within this regime $t_{\rm long}$ is too large for the entire process to be observable on human timescales, at least for the specific case of SgrA$^*$. From this perspective, it therefore appears that prospects for observing the BHSR-induced evolution of the shadow of SgrA$^*$ are significantly more favorable for the single-species rather than two-species case (particularly once care is given to the identification of the observationally relevant timescale, $t_{\rm short}$ rather than $t_{\rm long}$), contrary to the initial na\"{i}ve expectation laid out at the start of this work.

Remaining in the $0.05 \lesssim \beta \lesssim 0.5$ regime, instead of restricting ourselves to the case of SgrA$^*$ we can ask ourselves a different question: for a given value of $\beta$, what is the largest allowed value for the BH mass which still leads to an observable evolution timescale? For concreteness, we take $\beta=0.2$ which, as argued earlier, leads to comparable contributions from both species ($p_1 \sim p_2 \sim 0.5$). From Fig.~\ref{fig:densityPlot} (upper left panel), we see that the BH mass needs to be $\lesssim 4 \times 10^5M_{\odot}$ in order for the total evolution timescale to be observable ($t_{\rm long} \lesssim 50\,{\rm yrs}$).

For the same choice of parameters, the BH shadow size (upper right panel) changes by $\Delta R \sim {\cal O}(0.1)M$. How this number translates to an absolute angular scale depends on the distance to the BH: for instance, for a BH located $16.8\,{\rm Mpc}$ away from us (as is the case for M87$^*$), the BH shadow angular size (lower left panel) changes by $\Delta R_{\rm Abs} \sim {\cal O}(10^{-5})\,\mu{\rm as}$, which is way beyond the detection capabilities of the EHT or near-future upgrades thereof. For the previous figure to stand a chance to be detectable in future VLBI arrays, the distance to the BH would need to be ${\cal O}({\rm kpc})$ or smaller. However, there are no known BHs at ${\cal O}({\rm kpc})$ distance from us with mass in the ${\cal O}(10^5)M_{\odot}$ range: there are approximately 12 candidate BHs which are closer to us than SgrA$^*$ (including Cyg X-1, although only for some of these candidates has their BH nature been widely agreed upon), with masses in the ${\cal O}(10^{-1}-10)M_{\odot}$ range, hence too small to lead to observable changes in the angular size of the shadow. For the same choice of parameters, the change in the axis ratio (lower right panel) depends to very good approximation only on $\beta$ and not on $M_0$, but is undetectably small by about an order of magnitude, since $\Delta \chi \lesssim {\cal O}(0.1)$.

Looking at the bigger picture, let us return to the initial overarching question motivating our study: is it feasible to observe the BHSR-induced evolution of BH shadows in the not-too-distant future? A short answer is ``\textit{yes, in principle}'', and hinges upon \textit{a)} observationally favorable conditions, \textit{b)} small but non-negligible improvements in the angular resolution accessible to BH shadow imaging techniques (be these VLBI or other techniques, to be discussed soon), and \textit{c)} a careful definition of the observationally relevant evolution timescale.

For what concerns \textit{a)}, our analysis appears to suggest that the best observational prospects are found for SMBHs in the comparatively low-mass end of the window which are still sufficiently close to us, the archetype example being SgrA$^*$. For BHs of lower mass, the absolute change in the BH shadow angular size is too small and way below the resolution of future VLBI technology (unless the BH in question is extremely close to us, although no such candidate is known), whereas for BHs of higher mass the evolution timescale is too large. Moreover, contrary to our initial expectations, the best observational prospects for SgrA$^*$-like SMBHs actually occur for conditions as close to the single-species case as possible ($\beta \to 0$), as the presence of multiple species extracting the BH spin leads to a significant decrease in the superradiance rate, resulting in the final BH evolution being slower, rather than faster as initially na\"{i}vely expected.

Regarding \textit{b)}, even if the conditions set in \textit{a)} are met, reaching sub-$\mu{\rm as}$ angular resolutions with BH imaging techniques remains vital for the effects we have outlined to be observable. Such a resolution appears to be a factor of a few below the sensitivity achievable with a future upgrade to the EHT including additional telescopes in space~\cite{Fish:2019epg}. Another possibility is VLBI in the optical band, first suggested in Ref.~\cite{Tsupko:2019pzg}, for which however the presence of dust surrounding active galactic nuclei may be a significant limiting factor, as pointed out in Ref.~\cite{Vagnozzi:2020quf}. Moving away from VLBI techniques, and further down the timeline, an interesting possibility is that of using constellations of satellites to perform X-ray interferometry (XRI). By $\sim$2060, XRI may be able to achieve sub-$\mu{\rm as}$ angular resolution~\cite{Uttley:2019ngm}. It should be noted, however, that XRI is best suited for imaging SMBHs with optically thick disks: in this case, what is being observed (which we refer to as the ``XRI BH shadow'') is no longer the apparent image of the photon region, but the inner edge of the accretion disk. The fact that the location of the latter depends strongly on the BH spin may actually be an advantage, as the associated changes in the XRI BH shadow may be larger due to superradiant spin extraction. A full investigation goes beyond the scope of this paper, and is deferred to future work.

With regard to \textit{c)} we have argued in detail that, at least for the observationally more favorable single-species case (see point \textit{b)} above), $t_{\rm short}$ is a more significant definition of the relevant timescale than the previously adopted $t_{\rm long}$. For the two-species case, a more careful assessment is required when identifying the relevant timescale. While it is the case that in both the small- and large-$\beta$ limits ($\beta \lesssim 0.05$ and $\beta \gtrsim 0.5$ respectively) the evolution is effectively very close to that of a single-species system, an experiment operating over extremely long timescales (well above human ones) should eventually be able to observe the effects associated to the two successive decreases in the BH spin evolution due to the mass splitting, at least in the small-$\beta$ regime (see Fig.~\ref{fig:tshort3}). While an unambiguous observation of the imprint of a decrease in the BH spin on the BH shadow would be a breakthrough scientific discovery, the degeneracy between the effect being due to a genuine single species or two species within one of the regimes described above can only be discerned by observing across the complete timescale $t_{\rm long}$. However, we remark that for all relevant cases $t_{\rm long}$ is well above typical human timescales.

Given that SMBHs such as SgrA$^*$ have been around for an extremely long time, one might legitimately wonder why we should be so lucky as to observe the superradiance-induced evolution of the BH shadow precisely now (or in any case during our lifetime). In fact, if SMBHs can only go through a superradiant instability phase once during their existence, there is no good reason to expect the onset of the superradiant instability, and therefore the effects we have presented, to occur during our lifetime. Taking $t_{\rm BH}$ to be the typical lifetime of a SMBH, we can na\"{i}vely estimate the probability of observing the superradiance-induced shadow evolution now as being $p \sim t_{\rm short}/t_{\rm BH}$. For SgrA$^*$, which might have been around for billions of years, $t_{\rm BH} \sim {\cal O}(10^9)\,{\rm yrs}$ and therefore $p \sim {\cal O}(10^{-8})$. This na\"{i}ve estimate appears to suggest that observing the onset of superradiance during our lifetime would require nothing short of a coincidence.

However, in the presence of effects competing with superradiance such as gas accretion, the basic underlying assumption that a SMBH can only go through a superradiant instability phase once during its existence no longer holds. In fact, once a SMBH has spun down to the minimum spin (or in any case close enough to the minimum spin that the superradiance rate is negligible), it can be spun-up by accretion over a timescale $t_{\rm ACC}$, typically much longer than the superradiance timescale. Once the spin is sufficiently high, the superradiance process can start once more. In principle, modulo variations in the accretion rate over cosmic time, this cycle can eventually repeat several times (possibly in a quasi-periodic way).~\footnote{A clear visual representation of the impact of accretion in re-spinning a SMBH which has reached its minimum spin at the end of the superradiance process can be found in the upper panels of Fig.~5 of Ref.~\cite{Brito:2014wla} (compare the right and left panels).} This highlights the importance of considering the competing effect of gas accretion, even though its impact on our earlier results (and in particular on the estimation of $t_{\rm short}$) was nominally small.

For a SMBH of mass ${\cal O}(10^6)M_{\odot}$ such as SgrA$^*$, using Eq.~(\ref{eq:dotmacc}) we can estimate the typical timescale for accretion to operate and re-spin the SMBH as being $t_{\rm ACC} \sim 1000\,{\rm yrs}$. This completely changes the previous estimate of the probability of observing the superradiance-induced shadow evolution during our lifetime, which is now given by $p \sim t_{\rm short}/t_{\rm ACC} \sim 2\%$, a figure which is not tiny. Of course, this estimate relies on the relatively simple modelling of the effect of gas accretion we have adopted, following Ref.~\cite{Brito:2014wla}, whereas realistic astrophysical environments likely require a more sophisticated description, which may either enhance or decrease the probability of observing the superradiance-induced shadow evolution during our lifetime. Nonetheless, what is important is to note the key role of accretion in allowing the superradiant instability process to take place (thereby bringing all the associated observational effects we have discussed in this paper) more than once during the lifetime of a SMBH.

\section{Conclusions}
\label{sec:conclusions}

Over the past years, black holes have emerged as novel and powerful windows onto fundamental physics. In particular BH superradiance, the buildup and amplification of a bosonic cloud surrounding a BH at the expense of the BH's mass and angular momentum, turns BHs into an unique probe of new ultralight particles, which appear ubiquitously in well-motivated theoretical scenarios. At the same time, the first detection of the shadow of M87$^*$ by the EHT collaboration opens up the possibility for novel tests of fundamental physics in the strong-field regime. This work ties together BH shadows and superradiance, to further explore whether BH shadows can be used to test for the possible existence of ultralight particles: as the appearance of a BH shadow (imaged for instance through VLBI techniques) depends directly on the BH's mass and spin, the superradiance-induced evolution of these quantities is expected to lead to a time-evolution of the BH shadow. Earlier work in Refs.~\cite{Roy:2019esk,Creci:2020mfg} argued that the associated changes in the BH shadow are unlikely to be observable, either because they are too small or because they operate over timescales which are too long compared to typical human timescales.

In this work, we have revisited the issue of whether the superradiance-induced BH shadow evolution can be observationally detected. In particular, we have gone beyond the earlier works in Refs.~\cite{Roy:2019esk,Creci:2020mfg} in at least three respects. Firstly, we have performed a more realistic and consistent modelling of the superradiant evolution, including at the same time the competing effects of gas accretion and GW emission. Next, we have expanded the analysis to include two ultralight species, as a first step towards a multi-species analysis, given that several well-motivated scenarios extending the SM predict the existence of a plethora of ultralight species (as in the case of the string axiverse). Finally, we have pointed out important subtleties with regards to the observationally relevant definition of the BH shadow evolution timescale, which had earlier been missed.

Our findings indicate that, contrary to earlier results, the superradiance-induced evolution of BH shadows is potentially observable, while not currently, at least with near-future technology. An important aspect towards reaching this conclusion is a careful assessment of the observationally relevant evolution timescale: while the complete superradiant evolution occurs over a long timescale which exceeds typical human timescales (and which we referred to as $t_{\rm long}$), the most significant part of the evolution occurs over a shorter timescale which is about an order of magnitude lower (and which we referred to as $t_{\rm short}$). Under favorable parameter choices, the superradiant-induced BH shadow evolution may therefore be potentially observable on human timescales.

Contrary to our initial expectations, we have found that the inclusion of an additional ultralight species does not aid the observational prospects of the process, for which the most favorable conditions appear to occur as close to the single-species case as possible. Moreover, we have found the effects of gas accretion and GW emission to be negligible, confirming the goodness of previous simplified linearized analyses neglecting these effects. Overall, due to a trade-off between the (absolute) size of the associated changes in the BH shadow and the evolution timescale, we have identified the most promising systems for observing the superradiance-induced evolution of BH shadows to be SgrA$^*$-like systems, i.e.\ low-mass SMBHs situated sufficiently close to us.

Our results further strengthen the scientific case for imaging SgrA$^*$'s shadow, a goal which is anyhow within the upcoming plans of the EHT. However, the possibility of observing these effects hinges upon improvements in the angular resolution achievable by BH imaging techniques, which are required to reach sub-$\mu{\rm as}$ precision: while VLBI with additional telescopes in space or in the optical band do not appear promising in this sense, we have briefly commented upon X-ray interferometry (XRI) as a potentially promising technique, albeit on a much longer timescale ($\sim 2060$).

The results we have presented are certainly not the final word on the subject, as there are many interesting follow-up directions. It could be worth exploring the impact of having more than two ultralight species and investigating prospects of constraining the mass spectra of theoretical scenarios featuring multiple ultra-light particles (such as the string axiverse), studying whether superradiance-induced evolution of BH shadows can be used to test specific well-motivated dark matter and/or dark energy models, or examining whether considering higher excitation modes can improve our results. We remark that superradiance not only occurs for spinning BHs described within general relativity, but the phenomenon appears for any asymptotically-flat spacetime with an event horizon. Superradiant instabilities have been studied in theories of modified gravity (see e.g. Refs.~\cite{Koga:1994np,Khodadi:2020cht,Khodadi:2021owg,Kolyvaris:2018zxl,Zhang:2020sjh,Liu:2020evp,Rahmani:2020vvv}), where the resulting superradiant timescale can be significantly shorter, and thereby potentially leading to more optimistic results in the context of our work. For example, superradiance effects are strongly amplified for rotating dilatonic BHs in compactified higher dimensions~\cite{Koga:1994np}.

Worthy of further investigation are also more detailed studies on the complementarity between our results and other observational probes of ultralight particles such as cosmological~\cite{Baumann:2015rya,Baumann:2017gkg,CMB-S4:2016ple,SimonsObservatory:2018koc,SimonsObservatory:2019qwx}, astrophysical~\cite{Jain:2012tn,Giannotti:2015kwo,Croon:2020oga,Sakstein:2020axg,Caputo:2021eaa}, astronomical~\cite{Sun:2019ico,KumarPoddar:2020kdz,Tsai:2021irw,Benisty:2021cmq,Poddar:2021ose}, and laboratory tests~\cite{Wagner:2012ui,Caldwell:2016dcw,Burrage:2017qrf,Bloch:2020uzh,Vagnozzi:2021quy}. On the more technical side, we have made rather minimal assumptions on the ultralight (pseudo)scalar sector, but these could be relaxed in order to entertain a scenario which is theoretically (and ideally observationally) richer, featuring for instance self-interactions and kinetic mixing between different fields. Also worthy of much deeper investigations are future observational prospects of reaching sub-$\mu{\rm as}$ resolution with BH shadow imaging techniques, for instance through approaches such as XRI: while XRI is not expected to observe the gravitationally lensed image of the photon region, it might be much more sensitive to the effects associated to superradiant spin extraction on the ``XRI BH shadow''. While we leave a full study of these and other issues to follow-up projects, our work reinforces the message that BHs and their shadows are already at present remarkable probes of fundamental physics, including new ultralight particles.

\begin{acknowledgements}
\noindent We are grateful to Urjit A. Yajnik for useful comments and collaboration in the initial stages of this project. R.R.\ is supported by the Shanghai Government Scholarship (SGS). S.V.\ is supported by the Isaac Newton Trust and the Kavli Foundation through a Newton-Kavli Fellowship, and by a grant from the Foundation Blanceflor Boncompagni Ludovisi, n\'{e}e Bildt. S.V.\ acknowledges a College Research Associateship at Homerton College, University of Cambridge. L.V.\ acknowledges support from the European Union's Horizon 2020 research and innovation programme under the Marie Sk{\l}odowska-Curie grant agreement ``TALeNT'' No.~754496 (H2020-MSCA-COFUND-2016 FELLINI).
\end{acknowledgements}

\bibliographystyle{apsrev4-1}
\bibliography{BHsuperradiance}

\end{document}